\pdfoutput=1

\documentclass[11pt,twoside,a4paper,cmspaper,final,collab]{cms-tdr}
\def\svnVersion{463787P}\def\cmsCernNoTag{CERN-EP-2017-299}\def\cmsCernDate{\today}\def\cmsMessage{Published in the Journal of High Energy Physics as \href{http://dx.doi.org/10.1007/JHEP06(2018)027}{\doi{10.1007/JHEP06(2018)027.}}}

\begin{document}\cmsNoteHeader{EXO-16-051}

\hyphenation{had-ron-i-za-tion}
\hyphenation{cal-or-i-me-ter}
\hyphenation{de-vices}
\RCS$Revision: 462938 $
\RCS$HeadURL: svn+ssh://svn.cern.ch/reps/tdr2/papers/EXO-16-051/trunk/EXO-16-051.tex $
\RCS$Id: EXO-16-051.tex 462938 2018-06-02 02:31:36Z snarayan $

\newcommand{\Zvv}{\ensuremath{\PZ\to\nu\nu}}
\newcommand{\Zvvjets}{\ensuremath{\PZ(\nu\nu)\text{+jets}}}
\newcommand{\Wlvjets}{\ensuremath{\PW(\ell\nu)\text{+jets}}}
\newcommand{\phojets}{\ensuremath{\gamma\text{+jets}}}
\newcommand{\Vrm}{\ensuremath{\mathrm{V}}}
\newcommand{\Arm}{\ensuremath{\mathrm{A}}}
\providecommand{\NA}{\ensuremath{\text{---}}}
\ifthenelse{\boolean{cms@external}}
{\providecommand{\suppMaterial}{the supplemental material
  [URL will be inserted by publisher]}}
{\providecommand{\suppMaterial}{Appendix~\ref{sec:supp}}}
\ifthenelse{\boolean{cms@external}}
{\providecommand{\suppMaterialii}{the supplemental material}}
{\providecommand{\suppMaterialii}{Appendix~\ref{sec:supp}}}

\cmsNoteHeader{EXO-16-051}
\title{Search for dark matter in events with energetic, hadronically decaying top quarks and missing transverse momentum at $\sqrt{s}=13\TeV$}

\date{\today}

\abstract{
A search for dark matter is conducted in events with large missing transverse momentum and a hadronically decaying, Lorentz-boosted top quark.
This study is performed using proton-proton collisions at a center-of-mass energy of 13\TeV, in data recorded by the CMS detector in 2016 at the LHC, corresponding to an integrated luminosity of 36\fbinv.
New substructure techniques, including the novel use of energy correlation functions, are utilized to identify the decay products of the top quark.
With no significant deviations observed from predictions of the standard model, limits are placed on the production of new heavy bosons coupling to dark matter particles.
For a scenario with purely vector-like or purely axial-vector-like flavor changing neutral currents, mediator masses between 0.20 and 1.75\TeV are excluded at 95\% confidence level, given a sufficiently small dark matter mass.
Scalar resonances decaying into a top quark and a dark matter fermion are excluded for masses below 3.4\TeV, assuming a dark matter mass of 100\GeV.
}

\hypersetup{%
pdfauthor={CMS Collaboration},%
pdftitle={Search for dark matter in events with energetic, hadronically decaying top quarks and missing transverse momentum at sqrt(s)=13 TeV},%
pdfsubject={CMS},%
pdfkeywords={CMS, physics, dark matter, substructure}}

\maketitle

\section{Introduction}\label{sec:theory}
The existence of dark matter (DM) can be inferred through astrophysical observations of its gravitational interactions~\cite{dm1,dm2,dm3}.
The nature of DM has remained elusive, although it is widely believed that it may have a particle physics origin.
Multiple models of new physics predict the existence of weakly interacting, neutral, massive particles that provide excellent sources of DM candidates.
Searches for DM are often carried out through direct searches for interactions between cosmic DM particles and detectors (\eg, via nuclear recoil~\cite{Akerib:2012ys}), or for particles produced in the annihilation or decay of relic DM particles~\cite{ams}.
The CERN LHC presents a unique opportunity to produce DM particles as well as study them.
In this paper, we describe a search for events where DM particles are produced in association with a top quark (hereafter called ``monotop''), originally proposed in Ref.~\cite{Andrea:2011ws}.
The associated production of a top quark and invisible particles is heavily suppressed in the standard model (SM).
Therefore, this signature can be used to probe the production of DM particles via a flavor-violating mechanism, which most DM models do not consider~\cite{dm2}.
Searches for the monotop final state have been carried out by the CDF experiment~\cite{PhysRevLett.108.201802} at the Fermilab Tevatron, and by the CMS~\cite{PhysRevLett.114.101801} and ATLAS~\cite{EurPhysJC.75.79} experiments at the CERN LHC at $\sqrt{s} = 8\TeV$.
The present search utilizes $13\TeV$ data accumulated by the CMS experiment in 2016, corresponding to an integrated luminosity of 36\fbinv.
To improve the sensitivity of the analysis compared to previous work, we employ new techniques for the reconstruction and identification of highly Lorentz-boosted top quarks.

In this search, we consider events with a top quark that decays to a bottom quark and a W boson, where the W boson decays to two light quarks.
The three quarks evolve into jets of hadrons.
This decay channel has the largest branching fraction (67\%) and is fully reconstructable.
Jets from highly Lorentz-boosted top quarks are distinguished from other types of hadronic signatures by means of a novel jet substructure discriminant, described in Section~\ref{sec:toptag}.

\begin{figure}[hb]
\centering
\includegraphics[width=0.4\columnwidth]{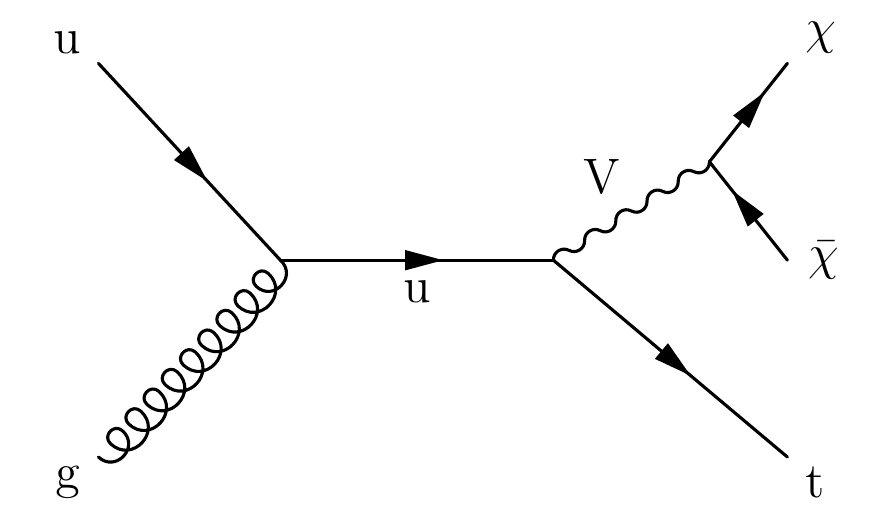}
\includegraphics[width=0.4\columnwidth]{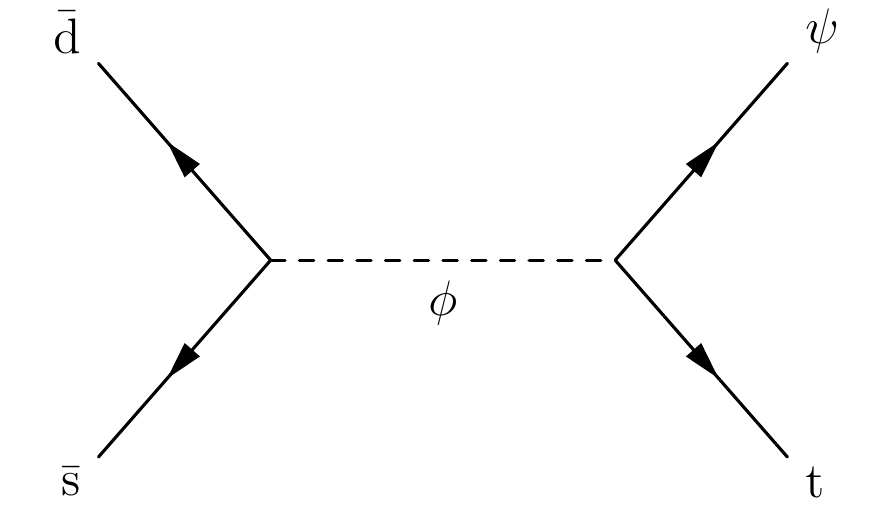}
\caption{Example Feynman diagrams of monotop production via a flavor-changing neutral current V (left) and a charged, heavy scalar resonance $\phi$ (right).}
\label{fig:diagrams}
\end{figure}

We interpret the results in terms of two monotop production mechanisms, example Feynman diagrams for which are shown in Fig.~\ref{fig:diagrams}.
One model involves a flavor-changing neutral current (FCNC), where a top quark is produced in association with a vector boson that has flavor-changing couplings to quarks and can decay to a pair of DM particles.
This is referred to in this paper as the ``nonresonant'' mode.
In a simplified model approach, the interaction terms of the effective Lagrangian~\cite{Agram:2013wda,Boucheneb2015,Andrea:2011ws} describing nonresonant monotop production are given by:
\begin{linenomath}
\begin{equation}
\mathcal{L}_\text{int}=  \Vrm_\mu  \overline\chi \gamma^\mu (  g^\Vrm_{\chi} + g^\Arm_{\chi} \gamma_5 ) \chi
                           + \overline{\mathrm{q}}_\mathrm{u} \gamma^\mu
                           ( g^\Vrm_\mathrm{u} + g^\Arm_\mathrm{u} \gamma_5 ) \mathrm{q}_\mathrm{u} \Vrm_\mu
                           + \overline{\mathrm{q}}_\mathrm{d} \gamma^\mu
                           (  g^\Vrm_\mathrm{d} + g^\Arm_\mathrm{d} \gamma_5 ) \mathrm{q}_\mathrm{d} \Vrm_\mu
                           + \text{h.c.},
    \label{eq:Lnonres}
\end{equation}
\end{linenomath}
where ``h.c.'' refers to the Hermitian conjugate of the preceding terms in the Lagrangian.
The heavy mediator is denoted $\Vrm$, and $\chi$ is the DM particle, assumed to be a Dirac fermion.
The couplings $g^\Vrm_{\chi}$ and $g^\Arm_{\chi}$ are respectively the vector- and axial vector-couplings between $\chi$ and $\Vrm$.
In the quark-$\Vrm$ interaction terms, it is understood that $\mathrm{q}_\mathrm{u}$ and $\mathrm{q}_\mathrm{d}$ represent three generations of up- and down-type quarks, respectively.
Correspondingly, $g^\Vrm_\mathrm{u}$ and $g^\Arm_\mathrm{u}$ are $3\times3$ flavor matrices that determine the vector- and axial vector-couplings between V and u, c, and top quarks.
It is through the off-diagonal elements of these matrices that monotop production becomes possible.
To preserve $\text{SU}(2)_L$ symmetry, analogous down-type couplings $g^\Vrm_\mathrm{d}$ and $g^\Arm_\mathrm{d}$ must be introduced, and the following must be satisfied~\cite{Andrea:2011ws}:
\begin{linenomath}
\begin{equation}
  g^\Vrm_\mathrm{u} - g^\Arm_\mathrm{u} = g^\Vrm_\mathrm{d} - g^\Arm_\mathrm{d}.
\end{equation}
\end{linenomath}
By choice, we assume $g^\Vrm_\mathrm{u} = g^\Vrm_\mathrm{d} \equiv g^\Vrm_\mathrm{q}$, and $g^\Arm_\mathrm{u} = g^\Arm_\mathrm{d} \equiv g^\Arm_\mathrm{q}$, both satisfying the above constraint.
Moreover, to focus specifically on monotop production, the only nonzero elements of $g^\Vrm_\mathrm{q}$ and $g^\Arm_\mathrm{q}$ are assumed to be those between the first and third generations.

The second model contains a colored, charged scalar $\phi$ that decays to a top quark and a DM fermion $\psi$~\cite{Boucheneb2015}.
In this ``resonant'' model the interaction terms of the effective Lagrangian are given by:
\begin{linenomath}
\begin{equation}
\mathcal{L}_\text{int} = \phi\overline{\mathrm{d}}_i^C[(a_\mathrm{q})^{ij}+(b_\mathrm{q})^{ij}\gamma^5]\mathrm{d}_j+\phi\overline{\mathrm{t}}[a_{\psi}+b_{\psi}\gamma^5]\psi+\text{h.c.}
\end{equation}
\end{linenomath}
The Lagrangian includes interactions between the scalar resonance $\phi$ and down-type quarks $\mathrm{d}_i$, controlled by the couplings $a_\mathrm{q}$ (scalar) and $b_\mathrm{q}$ (pseudoscalar).
Similarly, the couplings $a_\psi$ and $b_\psi$ allow for the decay of $\phi$ to a top quark and a DM fermion $\psi$.
We assume $a_\mathrm{q}=b_\mathrm{q}=0.1$ and $a_{\psi} = b_{\psi} = 0.2$.
A detailed motivation of these conventions is given in Ref.~\cite{Boucheneb2015}.
Signal model kinematic distributions are presented in Figures~\ifthenelse{\boolean{cms@external}}{1--2}{\ref{fig:lonlo}-\ref{fig:sigkin}} in \suppMaterial.

\section{The CMS detector, particle reconstruction, and event simulation}
\label{sec:cms}

The CMS detector, described in detail in Ref.~\cite{CMSdetector}, is a multipurpose apparatus designed to study high-transverse momentum ($\pt$) processes in proton-proton and heavy-ion collisions.
A superconducting solenoid occupies its central region, providing a magnetic field of 3.8\unit{T} parallel to the beam direction.
Charged particle trajectories are measured using silicon pixel and strip trackers that cover a pseudorapidity region of $\abs{\eta} < 2.5$.
A lead tungstate (PbWO$_4$) crystal electromagnetic calorimeter (ECAL) and a brass and scintillator hadron calorimeter (HCAL) surround the tracking volume and extend to $\abs{\eta} < 3$.
The steel and quartz-fiber forward Cherenkov hadron calorimeter extends the coverage to $\abs{\eta} < 5$.
The muon system consists of gas-ionization detectors embedded in the steel flux-return yoke outside the solenoid and covers $\abs{\eta} < 2.4$.
The return yoke carries a 2\unit{T} return field from the solenoid.
The first level of the CMS trigger system is designed to select events in less than 4\mus, using information from the calorimeters and muon detectors.
The high-level trigger-processor farm reduces the event rate to several hundred Hz.

The particle-flow (PF) event algorithm~\cite{Sirunyan:2017ulk} reconstructs and identifies each individual particle through an optimized combination of information from the different elements of the CMS detector.
The energy of a photon is obtained directly from the ECAL measurement, corrected for effects from neglecting signals close to the detector noise level (often termed zero-suppression).
The energy of an electron is determined from a combination of the electron momentum at the primary interaction vertex as determined by the tracker, the energy of the corresponding ECAL cluster, and the energy sum of all photons spatially compatible with originating from the electron track.
The energy of a muon is obtained from the curvature of the corresponding track.
The energy of a charged hadron is determined from a combination of its momentum measured in the tracker and the matching ECAL and HCAL energy deposits, corrected for zero-suppression effects and for the response function of the calorimeters to hadronic showers.
Finally, the energy of a neutral hadron is obtained from the corresponding corrected ECAL and HCAL energy.

The DM signal is generated by Monte Carlo (MC) simulation using \MGvATNLO v2.4.3~\cite{amcatnlo}.
Events for the nonresonant production are calculated at next-to-leading order (NLO) in quantum chromodynamics (QCD) perturbation theory.
Masses for the mediator V of $m_\Vrm >200\GeV$ are considered in order to provide an SM-like top quark width, i.e., to avoid decays of the top quark into a u quark plus an on-shell (for $m_\Vrm<m_\mathrm{t}$) or off-shell (for $m_\mathrm{t}\approx m_\Vrm-\Gamma_\Vrm$) mediator V, where $\Gamma_\Vrm$ is the width of V.
The resonant mode is generated at leading order (LO) accuracy.

To model the expectations from SM backgrounds, the \ttbar and single top quark backgrounds are generated at NLO in QCD using \POWHEG~v2~\cite{Nason:2004rx,Frixione:2007vw,Alioli:2010xd}.
Predictions for ZZ, WZ, or WW (i.e., diboson) production are obtained at LO with {\PYTHIA 8.205}~\cite{Sjostrand:2014zea}.
Events with multiple jets produced through the strong interaction (referred to as QCD multijet events) are simulated at LO using \MGvATNLO v2.3.3.
Simulated samples of Z+jets, W+jets, and $\gamma$+jets processes are generated at LO using \MGvATNLO v2.3.3, which matches jets from the matrix element calculations to parton shower jets using the MLM prescription~\cite{mlm}.
The samples are corrected by weighting the \pt of the respective boson with NLO QCD K-factors obtained from large samples of events generated with \MGvATNLO and the FxFx merging technique~\cite{fxfx}.
The samples are further corrected by applying NLO electroweak K-factors obtained from calculations~\cite{Kuhn:2005gv,Kallweit:2015fta,Kallweit:2015dum} that depend on boson \pt.

All samples produced using \MGvATNLO or {\POWHEG} are interfaced with {\PYTHIA 8.212} for parton showering, hadronization, and fragmentation, using the CUETP8M1~\cite{ue1,ue2} underlying-event tune.
The appropriate LO or NLO NNPDF3.0 sets ~\cite{Ball:2014uwa} are used for the parametrization of the parton distribution functions (PDF) required in all these simulations.
The propagation of all final state particles through the CMS detector are simulated with {\GEANTfour} \cite{geant4}.
To model the impact of particles from additional proton-proton interactions in an event (pileup), the number of simulated interactions is adjusted to match the distribution observed in the data~\cite{CMS-PAS-LUM-17-001}.
\section{Hadronically decaying top quark identification}
\label{sec:toptag}

For top quark $\pt>250\GeV$, the decay products are expected to be contained within a distance of $\Delta R = 1.5$ relative to the top quark, where $\Delta R = \sqrt{\smash[b]{(\Delta \eta)^2 + (\Delta\phi)^2}}$, and $\Delta\eta$ and $\Delta\phi$ are, respectively, the differences in pseudorapidities and azimuthal angles, where $\Delta\phi$ is measured in radians.
The final state particles of the hadronization of a light quark or gluon are reconstructed as a jet.
A standard jet-clustering algorithm at CMS is the anti-\kt algorithm~\cite{Cacciari:2008gp} with a distance parameter of 0.4 (AK4).
If a hadronically decaying top quark is highly Lorentz-boosted, reconstructing the three daughter quarks separately becomes difficult, as the resulting jets tend to overlap in the detector.
Accordingly, to identify such signatures, we define CA15 jets as objects that are clustered from PF candidates using the Cambridge--Aachen algorithm~\cite{cajets} with a distance parameter of 1.5.
To reduce the impact of particles arising from pileup, weights calculated with the pileup per particle identification (PUPPI) algorithm~\cite{puppi} are applied to the PF candidates.
Calibrations derived from data are then applied to correct the absolute scale of the jet energy~\cite{jec}.
The CA15 jets must pass the selection criteria $\pt>250\GeV$ and $\abs{\eta}<2.4$.
To be identified as arising from top quark decays, jets must have a mass within a specified interval containing the top quark mass, have a high likelihood of containing a bottom quark, and exhibit certain substructure characteristics.
Such jets are referred to as ``t-tagged'' jets hereafter.

The ``soft drop'' (SD)~\cite{msd} grooming method is used to remove soft and wide-angle radiation produced within jets through initial state radiation or through the underlying event.
Removing such radiation, the SD algorithm defines a subset of the CA15 jet's constituents, which are further grouped into subjets of the CA15 jet.
The grooming is done using the SD parameters $z_\text{cut}=0.15$ and $\beta=1$ (for their definition, see Ref.~\cite{msd}), chosen to optimize the resolution in the mass of the groomed jet  $m_\text{SD}$.
Hereafter, when the SD algorithm is referred to, these parameters are used.
We require t-tagged jets to satisfy $110 < m_\mathrm{SD} < 210\GeV$ to be compatible with the expectations of a top quark.

To identify the b quark in the CA15 jet expected from a top quark decay, we use the combined secondary vertex (CSVv2) algorithm~\cite{btag1,btag2}.
The b tagging criterion is then defined by requiring at least one subjet to have a CSVv2 score higher than a specified threshold.
The chosen threshold corresponds to correctly identifying a bottom jet with a probability of 80\%, and misidentifying a light-flavor jet with a probability of 10\%.

\subsection{Substructure}
\label{sec:bdt}

Three classes of substructure observables are employed to distinguish top quark jets from the hadronization products of single light quarks or gluons (hereafter referred to as ``q/g jets'').
These observables serve as inputs to a boosted decision tree (BDT)~\cite{tmva}, which is used as the final discriminator.

The $N$-subjettiness variable ($\tau_N$)~\cite{nsub} tests the compatibility of a jet with the hypothesis that it is composed of $N$ prongs.
For top quark decays, a three-pronged topology is expected, while q/g jets may have fewer prongs.
This makes the ratio $\tau_3/\tau_2$ a robust variable for top quark identification.
In this study, the $N$-subjettiness is computed after jet constituents have been removed using SD grooming, which reduces the \pt- and mass-dependence of $\tau_3/\tau_2$.

The \textsc{HEPTopTaggerV2} uses the mass drop and filtering algorithms~\cite{htt} to construct subjets within the CA15 jet.
The algorithm then chooses the three subjets that are most compatible with top quark decay kinematics.
The \textsc{HEPTopTaggerV2} defines a discriminating variable $f_\text{rec}$, which quantifies the difference between the reconstructed W boson and top quark masses and their expected values:
\begin{linenomath}
\begin{equation}
  f_\text{rec} = \min_{i,j} \left|\frac{m_{ij}/m_{123}}{m_\text{W}/m_\text{t}} - 1\right|,
\end{equation}
\end{linenomath}
where $i,j$ range over the three chosen subjets, $m_{ij}$ is the mass of subjets $i$ and $j
$, and $m_{123}$ is the mass of all three subjets.

Finally, energy correlation functions (ECF) $_{a}e^{(\alpha)}_N$ are considered, which are sensitive to correlations among the constituents of the jet~\cite{ecf0,ecf}.
They are $N$-point correlation functions of the constituents' momenta, weighted by the angular separation of the constituents in $\eta$ and $\phi$.
For a jet containing $N_{p}$ particles, an ECF is defined as
\begin{linenomath}
\begin{align}
  _{a}e^{(\alpha)}_N = & \sum_{1\leq i_1 < i_2 < \cdots < i_N \leq N_p}  \left[\prod_{1\leq k \leq N} \frac{\pt^{i_k}}{\pt^{J}}\right] \, \prod_{m=1}^a \left[\min^{(m)} \left\{\Delta R_{i_{j},i_{k}} \Big| {1 \leq j<k \leq N} \right\}\right]^\alpha,
  \label{eq:ecfdef}
\end{align}
\end{linenomath}
where $i_1,\dots,i_N$ range over the jet constituents.
The symbols $\pt^J$ and $\pt^{i_k}$ are, respectively, the \pt of the jet and the constituent $i_k$.
The notation $\min^{(m)}X$ refers to the $m$th smallest element of the set $X$.
We denote the distance $\Delta R$  between constituents $i_j$ and $i_k$ as $\Delta R_{i_j,i_k}$.
The parameters $N$ and $a$ must be positive integers, and $\alpha$ must be positive.

Discriminating substructure variables are constructed using ratios of these functions:
\begin{linenomath}
\begin{equation}
  \frac{_ae_N^{(\alpha)}}{\left(_be_M^{(\beta)}\right)^x}, \text{ where $M\leq N$ and $x= \frac{a\alpha}{b\beta}$}.
  \label{eq:ecf}
\end{equation}
\end{linenomath}
In Eq.~(\ref{eq:ecf}), the six adjustable parameters are $N$, $a$, $\alpha$, $M$, $b$, and $\beta$.
The value of $x$ is chosen to make the ratio dimensionless.
As with $N$-subjettiness, SD grooming is applied to the jet prior to computing the ECFs.

\begin{figure}[htbp]
  \centering
  \includegraphics[width=0.45\textwidth]{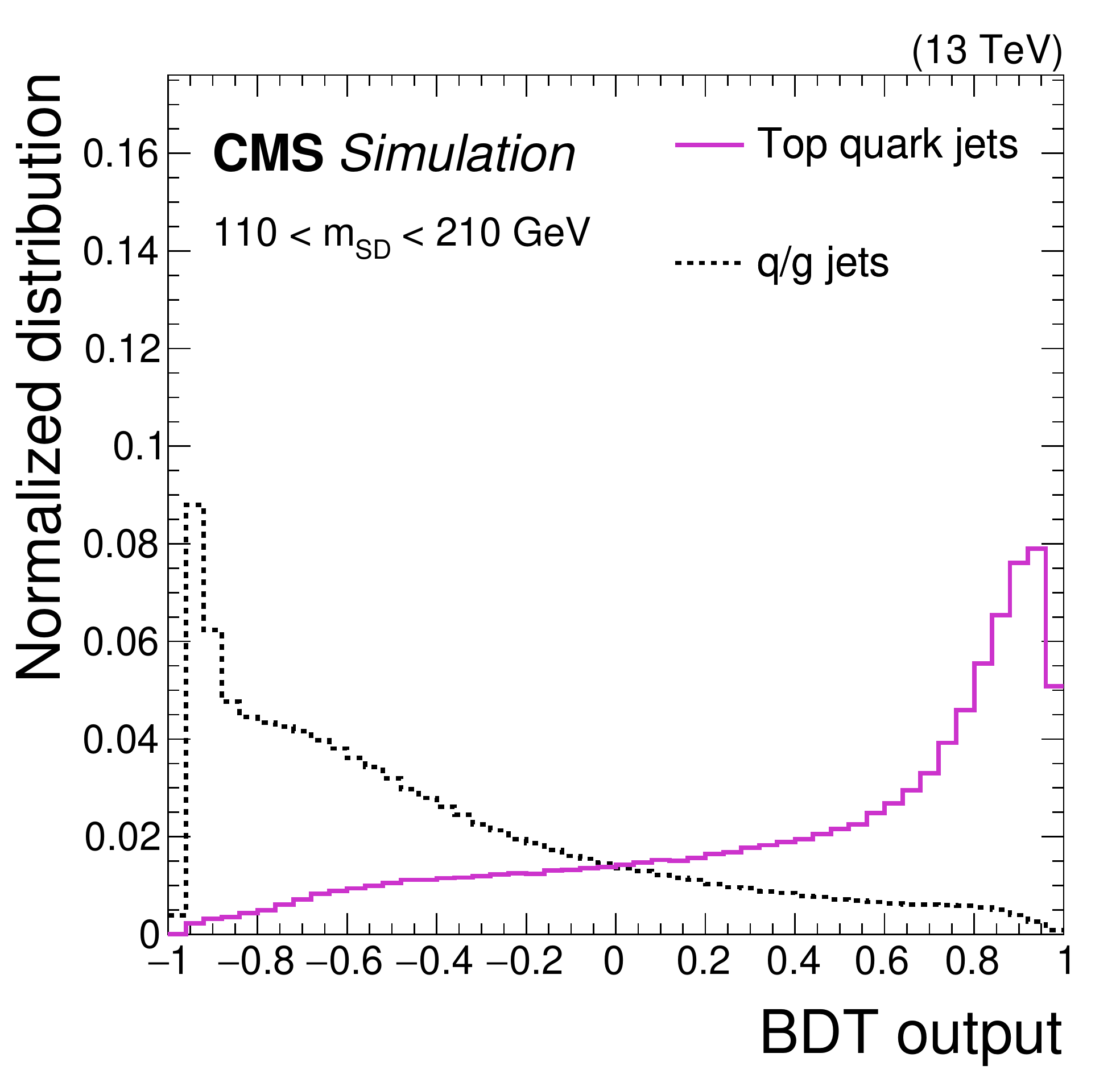}
  \includegraphics[width=0.45\textwidth]{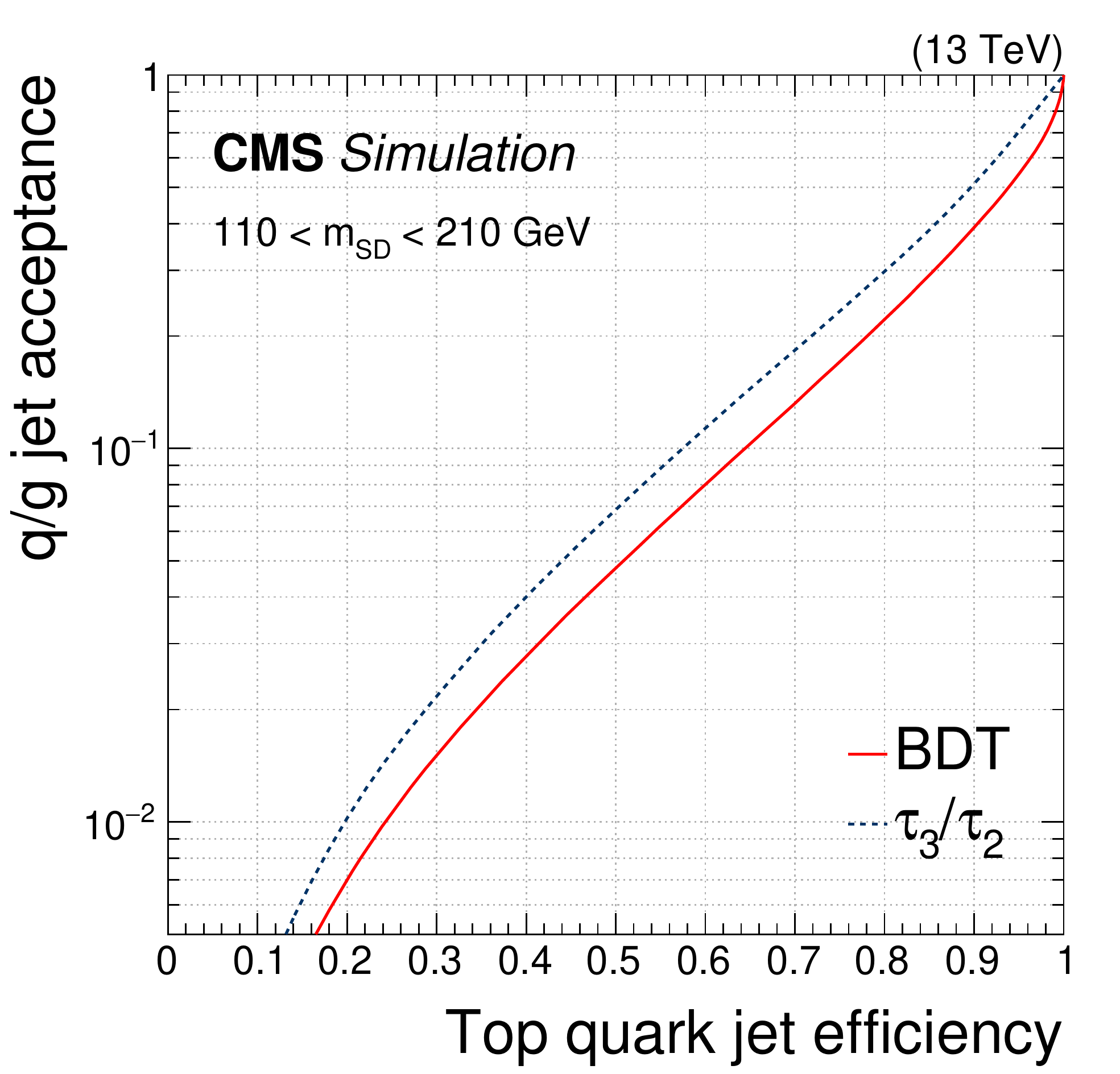}
  \caption{Performance of BDT tagging of top quark and q/g jets. The left figure shows the BDT output in both types of jets. The right figure shows the rate of misidentifying a q/g jet as a function of the efficiency of selecting top jets. In both figures, the \pt spectra of jets are weighted to be uniform, and the $m_\mathrm{SD}$ is required to be in the range of 110--210\GeV.}
  \label{fig:toptag_mc}
\end{figure}

\begin{figure}[htbp]
  \centering
  \includegraphics[width=0.45\textwidth]{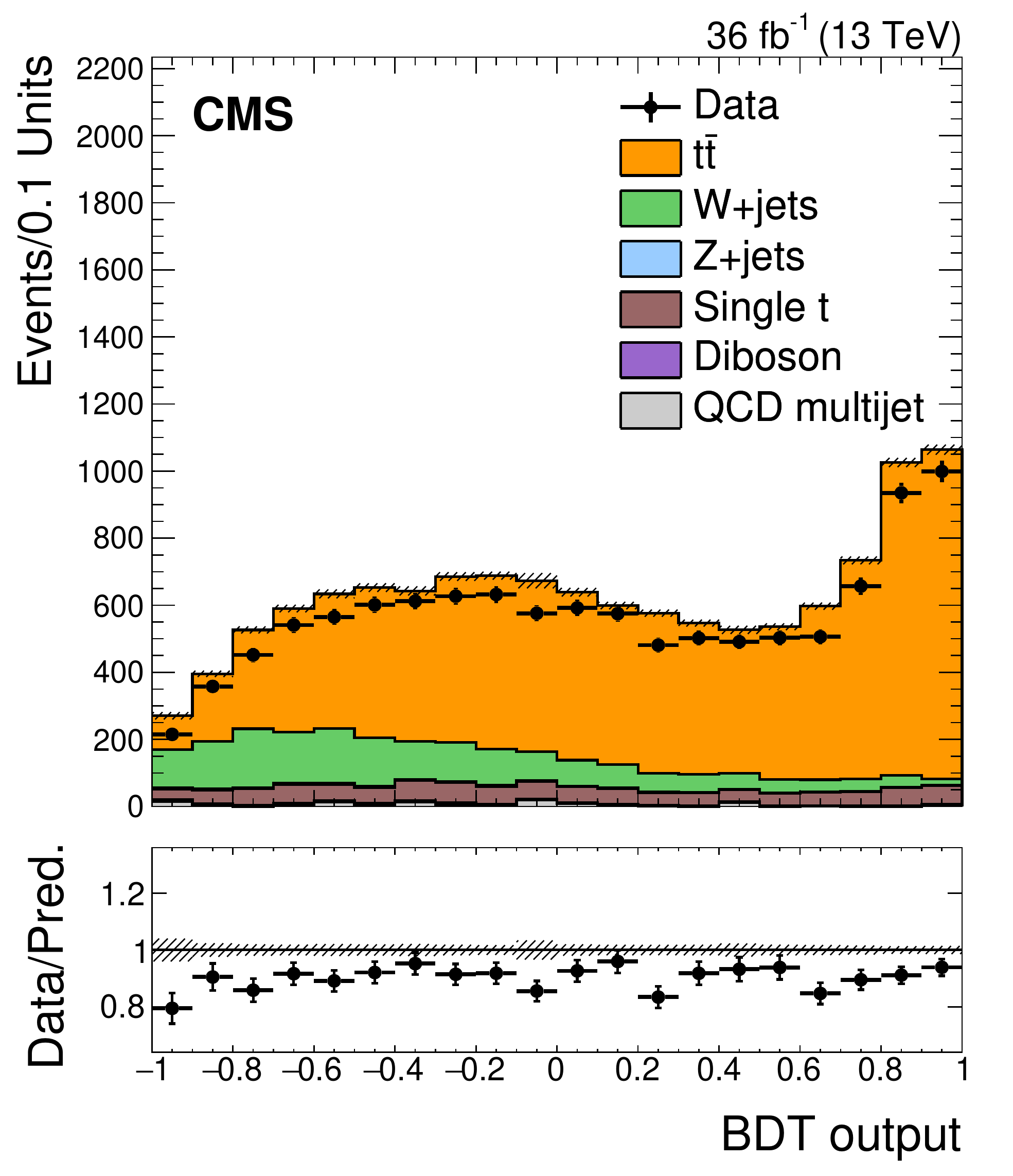}
  \includegraphics[width=0.45\textwidth]{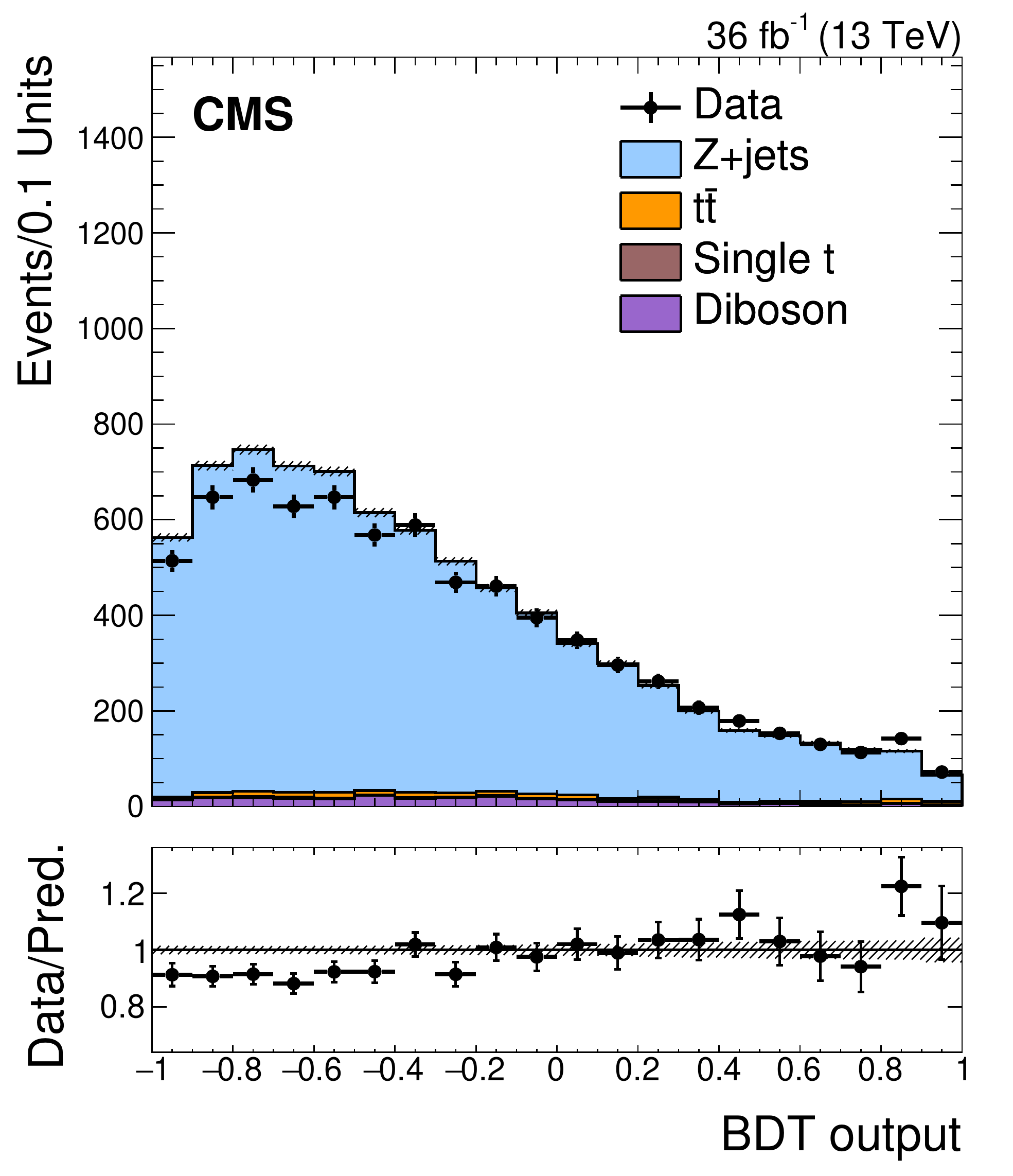} \\
  \includegraphics[width=0.45\textwidth]{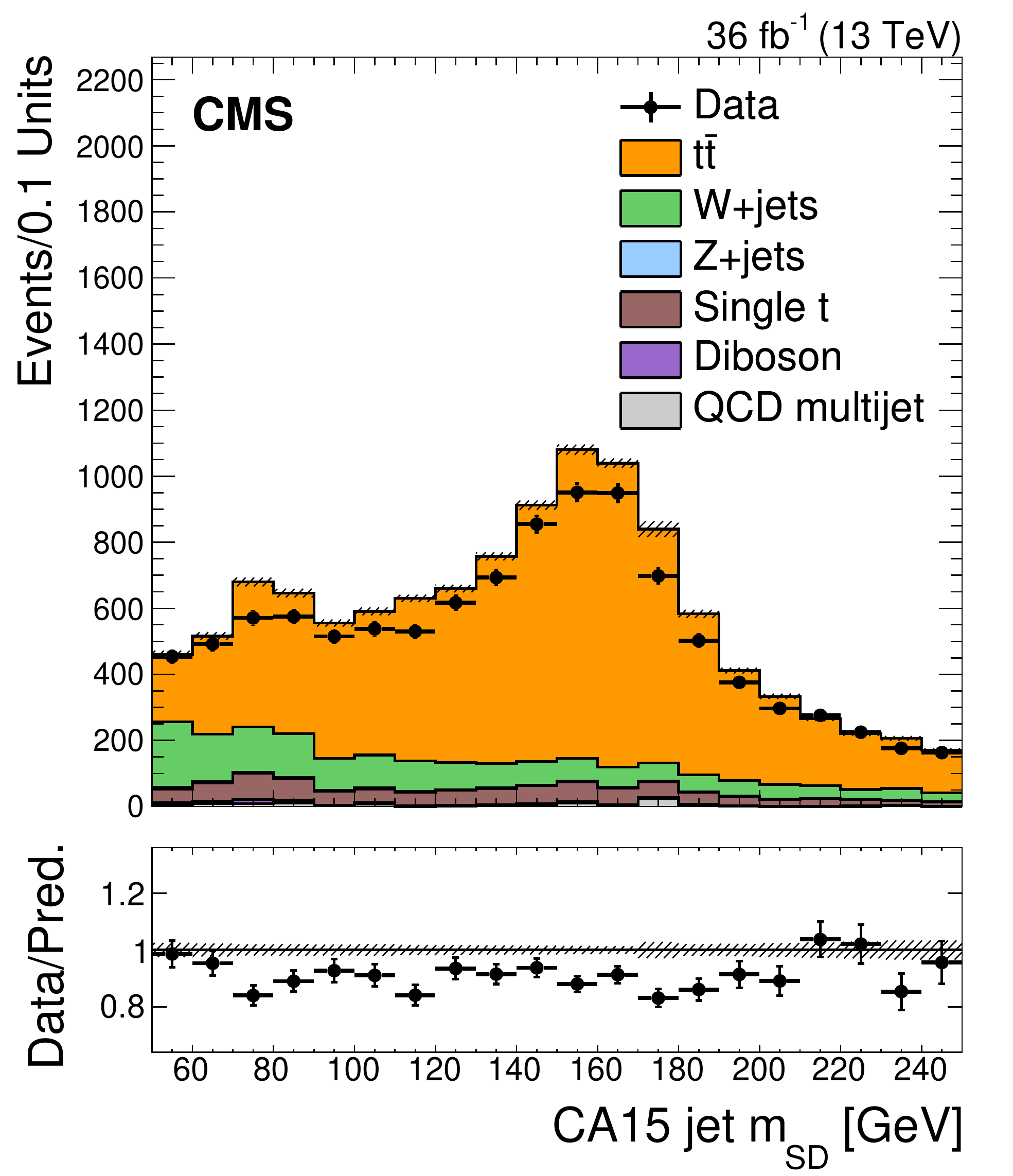}
  \includegraphics[width=0.45\textwidth]{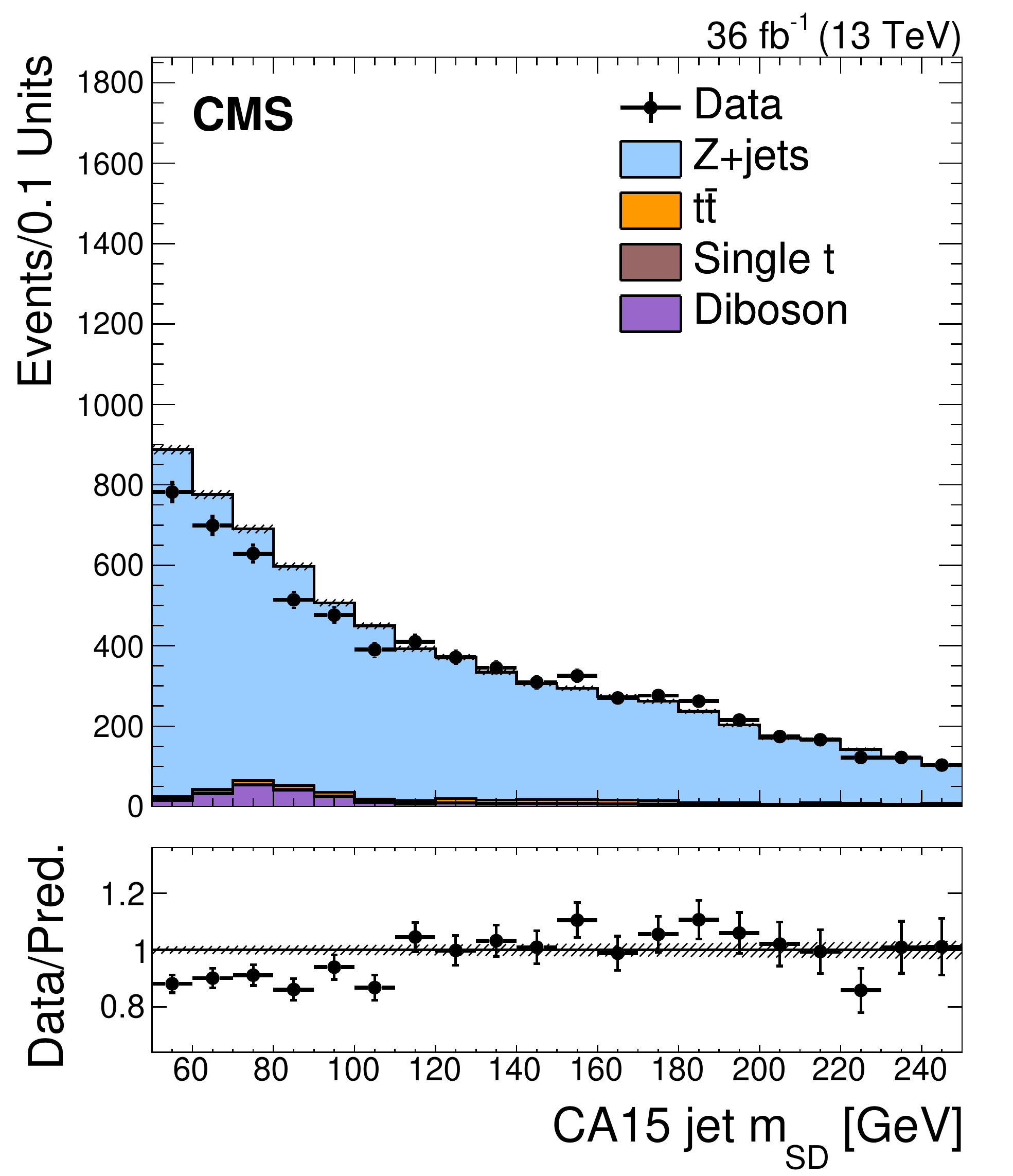}
  \caption{Comparison of the BDT response (upper) and $m_\mathrm{SD}$ (lower) in data and in simulation, in samples enriched in top-quark jets (left) and q/g jets (right). The lower panel of each plot shows the ratio of the observed data to the SM prediction in each bin. The shaded bands represent the statistical uncertainties in the simulation.}
  \label{fig:toptag_data}
\end{figure}

The following 11 ratios of ECFs are found useful for discriminating top quark jets from q/g jets:
\begin{linenomath}
\begin{equation}\begin{aligned}
      \frac{_1e_2^{(2)}}{\Bigl(_1e_2^{(1)}\Bigr)^2},
     \frac{_1e_3^{(4)}}{_2e_3^{(2)}},
     \frac{_3e_3^{(1)}}{\Bigl(_1e_3^{(4)}\Bigr)^{3/4}},
     &\frac{_3e_3^{(1)}}{\Bigl(_2e_3^{(2)}\Bigr)^{3/4}},
     \frac{_3e_3^{(2)}}{\Bigl(_3e_3^{(4)}\Bigr)^{1/2}},
     \\
      \frac{_1e_4^{(4)}}{\Bigl(_1e_3^{(2)}\Bigr)^{2}},
     \frac{_1e_4^{(2)}}{\Bigl(_1e_3^{(1)}\Bigr)^2},
     \frac{_2e_4^{(1/2)}}{\Bigl(_1e_3^{(1/2)}\Bigr)^{2}},
     &\frac{_2e_4^{(1)}}{\Bigl(_1e_3^{(1)}\Bigr)^{2}},
      \frac{_2e_4^{(1)}}{\Bigl(_2e_3^{(1/2)}\Bigr)^{2}},
       \frac{_2e_4^{(2)}}{\Bigl(_1e_3^{(2)}\Bigr)^{2}}.
      \label{eq:goodecfs}
\end{aligned}\end{equation}
\end{linenomath}
The final tagger is constructed by training a BDT using these thirteen variables ($\tau_3/\tau_2$, $f_\text{rec}$, and the ECF ratios) as inputs.
Figure~\ref{fig:toptag_mc} shows the BDT response and its performance in discriminating top quark jets from q/g jets.
At 50\% signal efficiency, the BDT background acceptance is 4.7\%, compared to 6.9\% for groomed $\tau_3/\tau_2$, which is commonly used for t tagging.
The distributions in BDT output and $m_\mathrm{SD}$ in MC and data are shown in Fig.~\ref{fig:toptag_data}, using control data enriched either in genuine top quark jets from \ttbar production or in q/g jets.
The selection of these control data is described in Section~\ref{sec:crs}.
In all distributions, a slight disagreement between data and simulation is observed.
This is accounted for by the use of data-driven estimates and scale factors, as described in Section~\ref{sec:signalest}.

\section{Event selection}\label{sec:selection}

\subsection{Signal topology selection} \label{sec:sigsel}

To search for monotop production, events are selected with two characteristic signatures: a large missing transverse momentum arising from DM candidates and a high-\pt CA15 jet from the decay of a top quark.
Events in the signal region (SR) are selected by a logical ``or'' of triggers with different minimum thresholds (90, 100, 110, or 120\GeV) for both $p_\mathrm{T,trig}^\text{miss}$ and $H_{\mathrm{T,trig}}^{\text{miss}}$.
In the trigger, $p_\mathrm{T,trig}^\text{miss}$ is defined by the magnitude of the vectorial \pt sum of all PF particles at the trigger level, and $H_{\mathrm{T,trig}}^{\text{miss}}$ by the magnitude of the similar sum of all AK4 jets with $\pt>20\GeV$ and $|\eta|<5.2$.
Muons are not included in these calculations.
Additional requirements are imposed on the energy depositions of the jets used to compute $H_{\mathrm{T,trig}}^{\text{miss}}$ to remove events resulting from instrumental effects.

In addition to CA15 jets, this search also utilizes jets clustered using the AK4 algorithm.
These will hereafter be referred to as ``AK4 jets'' and must have $\pt >30\GeV$ and satisfy $\abs{\eta}<4.7$.
The momenta of AK4 jets are corrected to account for mismeasurement of jet energy and for discrepancies between data and simulation~\cite{jec}.

The main observable in this analysis is \ptmiss, defined as the magnitude of the sum $\ptvecmiss$ of $\pt$ vectors of all final state particles reconstructed using the PF algorithm.
Corrections to the momenta of AK4 jets reconstructed in the event are propagated to the \ptmiss calculation.
A selected event is required to have $\ptmiss>250\GeV$.
The contribution from events with a large misreconstructed   \ptmiss value is reduced by removing events with beam halo particles, noise, or misreconstructed tracks.
For events passing the analysis selection, the efficiency of the triggers is found to be greater than 99\%.

To search for events with one hadronically decaying top quark and large $\ptmiss$, we require the presence of exactly one CA15 jet in the event.
The CA15 jet must pass the mass and b tagging requirements described in Section~\ref{sec:toptag}.
To account for discrepancies in b tagging between data and simulation, additional corrections are applied to simulated events.
The BDT described in Section~\ref{sec:toptag} is used to split the SR into two categories.
In the less restricted or ``loose'' category, the CA15 jet is required to have a BDT score greater than 0.1 and less than 0.45, while the ``tight'' category requires a minimum BDT score of 0.45.
These values were chosen to optimize the sensitivity of the search.

\subsection{Background rejection}

Monotop events with hadronically decaying top quarks are characterized by the signatures described in Section~\ref{sec:sigsel}.
Several SM processes can mimic these characteristics.
Events involving pair production of top quarks, in which one top quark decays to $\ell\nu\mathrm{b}$ and the other to $\mathrm{q}\overline{\mathrm{q}}'\mathrm{b}$, can have large \ptmiss and a CA15 jet.
Likewise, events with $\PW\to\ell\nu$ and $\Z\to\nu\nu$ can be characterized by large \ptmiss, and jets produced in association with the vector bosons can pass the t tagging selection.

To suppress these and other backgrounds, events are vetoed if they contain at least one well-identified and isolated electron, muon, tau lepton, or photon, passing the criteria described in the following paragraphs.

An electron or muon must have $\pt>10\GeV$.
In the case of electrons, additional criteria are imposed on the ECAL energy deposition, based on the distribution of energy in the shower and the presence of a nearby track~\cite{elid}.
To define an isolated electron, we compute the sum of the energies of the PF particles (charged and neutral hadrons and photons) within a cone of $\Delta R<0.3$ around the electron direction.
If this sum is less than 17.5\%\,(15.9\%) of the electron energy for electrons with $\abs{\eta}<1.479$ ($1.479<\abs{\eta}<2.5$), the electron is considered isolated.
In the case of muons, a track must be consistent with the energy depositions in the muon detectors.
An isolated muon is defined by setting an energy fraction ceiling of 20\% in a cone of $\Delta R<0.4$.
The tau leptons that decay to hadrons plus $\nu_\tau$ are required to have $\pt>18\GeV$ and are identified from jets that contain a subset of particles with a mass consistent with the decay products of a hadronically decaying tau lepton.
An additional set of identification and isolation criteria is applied to tau lepton candidates~\cite{Sirunyan:2017ulk}.
Photons must have $\pt>15\GeV$ and satisfy criteria on the distribution of energy depositions in the ECAL, to distinguish them from electrons or jets.
Furthermore, to avoid misidentifying an electron as a photon, the ECAL deposition of a photon candidate must not be near a track.

We define an isolated jet to be an AK4 jet that has $\Delta R >1.5$ relative to the CA15 jet.
Since isolated jets are only used to identify b jets, an isolated jet is further required to satisfy $\abs{\eta}<2.4$ and to lie within the tracker acceptance.
To reduce the \ttbar background, an event is rejected if there is an isolated jet that is likely to arise from the hadronization of a bottom quark.
The b jets are identified using the same CSVv2 algorithm and working point used to identify b quarks inside a CA15 jet.
As in the case of tagging CA15 jets, simulated events are corrected for discrepancies in the modeling of isolated jet b tagging.
To reduce the background from QCD multijet events in which large \ptmiss arises from the mismeasurement of jet momenta, the minimum azimuthal angle between the $\ptvecmiss$ direction and any AK4 jet has to be larger than 0.5\unit{rad}.

\section{Signal estimation} \label{sec:signalest}

A fit to the \ptmiss distribution in the SR is performed to search for the DM signal.
After applying the selection described in Section~\ref{sec:selection}, the dominant predicted backgrounds are \ttbar, \Zvvjets, and \Wlvjets.
The contributions from these SM processes are estimated using constraints from a simultaneous fit of seven control regions (CR), to be introduced in Section~\ref{sec:crs}.
The CRs are designed to target dimuon, dielectron, single-photon, single-muon, or single-electron events, with requirements on the substructure and the mass of the CA15 jet that are the same as in the SR.

In the CRs, the distribution of the backgrounds in \pt of recoiling jets ($\pt^\text{recoil}$) is used to model the \ptmiss distribution in the SRs.
The recoil $\pt^\text{recoil}$ is defined by removing leptons or photons (depending on the CR) from the \ptmiss calculation.
The primary backgrounds in the SR are constrained by defining transfer factors from the CRs to the SR in bins of $\pt^\text{recoil}$.
Additional information on the transfer factors and their theoretical and experimental uncertainties is given in Sections~\ref{sec:rfactor}~and~\ref{sec:unc}.
Each CR is split into loose and tight categories, using the same BDT criteria as the SR categories.
Each loose (tight) CR is used to constrain the target background only in the loose (tight) category of the SR.
Single top quark, diboson, and QCD multijet backgrounds are not constrained by the CR fit and are estimated using MC simulation.

A binned likelihood fit is performed simultaneously to the $\pt^\text{recoil}$ distributions in all signal and control regions.
The predictions from the CRs are translated to the SR through transfer factors that correlate corresponding bins across all regions.
These transfer factors can vary within their uncertainties, as described in Section~\ref{sec:unc}.

\subsection{Control regions} \label{sec:crs}

To estimate the contribution from $\Zvvjets$ in the SR, we use CRs enriched in dimuon, dielectron, and photon events.

Dimuon events are selected employing the same $p_\mathrm{T,trig}^\text{miss}$ triggers used in the SR, since these triggers do not include muons in the  $p_\mathrm{T,trig}^\text{miss}$ calculation.
Events are required to have two well-identified oppositely charged muons that form an invariant mass between 60 and 120\GeV.
At least one of the two muons must have $\pt > 20\GeV$ and pass tight identification and isolation requirements.
Events in the dimuon region must also pass almost all of the other selection requirements imposed on the events in the SR, wherein $\pt^\text{recoil}$ is substituted for \ptmiss.
To increase the number of events in the dimuon CR, the requirement for having a CA15 jet b tag is not imposed.

Dielectron events are selected using single-electron triggers, which have a \pt threshold of 27\GeV.
Two well-identified oppositely charged electrons are required, and they must form an invariant mass between 60 and 120\GeV.
To reach plateau efficiency with respect to the electron \pt, at least one of the two electrons must have $\pt>40\GeV$ and satisfy tight identification and isolation requirements.
All selection criteria applied in the dimuon CR are also applied in the dielectron CR.

The \phojets~control sample is constructed using events with at least one high-\pt photon.
A single-photon trigger with a \pt threshold of 165\GeV is used to record these events.
The event selection requires the photon to have a \pt  greater than 175\GeV in order to ensure that the trigger is fully efficient.
The photon candidate is required to pass identification and isolation criteria, and must be reconstructed in the ECAL barrel ($\abs{\eta} < 1.44$) to obtain a purity of 95\% ~\cite{cmsphotonid}.
As in the dilepton regions, the CA15 jet b tag requirement is not applied in the photon region.

Background events can enter the signal selection because of the loss of a single lepton, primarily from $\Wlvjets$ and lepton+jets \ttbar events.
To estimate these backgrounds, four single lepton control samples are used, defined by selecting electrons or muons and by requiring or vetoing b-tagged jets.
The b-tagged single lepton CRs are enhanced in \ttbar events, while the b-vetoed single lepton CRs target $\Wlvjets$ events.

Single-muon events are selected using the $p_\text{T,trig}^\text{miss}$ trigger.
The muon candidate in these events is required to have $\pt>20\GeV$, and pass tight identification and isolation requirements.
With the exception of b tagging, all other selection requirements used for signal events are imposed, using $\pt^\text{recoil}$ instead of \ptmiss.
In addition, to suppress QCD multijet events in which a jet passes the muon identification criteria and the \ptmiss is mismeasured, the transverse mass ($m_\mathrm{T}$) is required to be less than 160\GeV, where $m_\mathrm{T} = \sqrt{\smash[b]{2\ptmiss\pt^\ell(1-\cos\Delta\phi(\ptvecmiss,\vec{p}_\mathrm{T}^\ell))}}$.
In the b-tagged single-muon CR, we require the CA15 jet to be b-tagged as in the SR, and we further require exactly one b-tagged isolated jet.
In the b-vetoed single-muon CR, the b tagging requirements are reversed, so that the CA15 jet is not b-tagged and there are no b-tagged isolated jets.

The single-electron CRs are defined in a fashion similar to the single-muon CRs.
Events are selected using the single-electron trigger, and the \pt of the electron is required to be greater than $40\GeV$.
An additional requirement of $\ptmiss > 50\GeV$ is imposed on single-electron events to suppress the multijet background.

\begin{table}
\centering
    \topcaption{Summary of the selection criteria used in the SR and CRs. Symbols $\{\mathrm{b}\}$ and $\{\ell\}$ refer to cases where the b quark or lepton are not identified. The symbols $N_\Pe$, $N_\mu$,  and $N_\gamma$ refer to the number of selected electrons, muons, and photons, respectively. The number of b-tagged isolated jets is denoted with $N_{\text{b-tag}}^\text{iso}$.}
  \begin{tabular}{llccccc}
    \hline
    Region & Primary backgrounds & $N_\Pe$ & $N_\mu$  & $N_\gamma$ & $N_{\text{b-tag}}^\text{iso}$ & CA15 jet b-tag \\
    \hline
    & $\PW\to\{\ell\}\nu$, & & & & & \\
    Signal & $\Z\to\nu\nu$, & 0 & 0 & 0 & 0 & 1 \\
     & $\ttbar\to \{\mathrm{b}\}\mathrm{qq}'+\mathrm{b}\{\ell\}\nu$ & & & & & \\[1.8ex]
    \hline
     & Targeted contributions &  &  & &  &  \\
    \hline
    Single-e (b-tagged) & $\ttbar\to \mathrm{bqq}'+\mathrm{b}\Pe\nu$  & 1 & 0 & 0 & 1 & 1 \\
    Single-$\mu$ (b-tagged) & $\ttbar\to \mathrm{bqq}'+\mathrm{b}\mu\nu$  & 0 & 1 & 0 & 1 & 1 \\
    Single-e (b-vetoed) & $\PW\to \Pe\nu$ & 1 & 0 & 0 & 0 & 0 \\
    Single-$\mu$ (b-vetoed) & $\PW\to\mu\nu$ & 0 & 1 & 0 & 0 & 0 \\
    Dielectron & $\Z\to\Pe\Pe$ & 2 & 0 & 0 & 0 & \NA \\
    Dimuon & $\Z\to\mu\mu$ & 0 & 2 & 0 & 0 & \NA \\
    Photon & $\gamma$ & 0& 0 & 1 & 0 & \NA \\
    \hline
  \end{tabular}
  \label{tab:summary}
\end{table}

A summary of the selection criteria for the SR and for all of the CRs is given in Table~\ref{tab:summary}.

To account for discrepancies between data and simulation in efficiencies for identifying electrons, muons, and photons, correction factors are applied to simulated events in CRs where they are selected.

\subsection{Transfer factors}
\label{sec:rfactor}

The dominant SM process in each CR is used to estimate at least one background in the SR.
Each constraint is encoded through a transfer factor $T$, which is the ratio of the predicted yield of the targeted process in the SR and its predicted yield in the CR.
This factor is defined as a function of $\pt^\text{recoil}$ and is estimated using simulation.
If the CR $X$ is used to estimate the process $Y$ in the SR, then the number of events predicted in bin $i$ of the CR is $N^{X}_i = {\mu^Y_i}/{T^X_i}$, where $\mu^Y_i$ is the free parameter of the likelihood representing the number of events from process $Y$ observed in bin $i$ of the SR.

The \ttbar and W+jets backgrounds in the SR are estimated using data in the corresponding subsample of the single lepton CRs.
Transfer factors ($T^{\mathrm{b}\ell}$ and $T^{\ell}$) are obtained from simulations that take into account the effect of lepton acceptances and efficiencies, the b tagging efficiency, and, for the single-electron control sample, the additional \ptmiss requirement.
These transfer factors explicitly include hadronically decaying $\tau$ leptons that fail the identification criteria, which account for roughly 20\%--80\% of the total W+jets background in the high-recoil region.
Because of a large \ttbar contamination in the tight W+jets CR, an additional transfer factor is imposed between the \ttbar predictions in the b-tagged and b-vetoed single lepton CRs.
This provides an estimate of the \ttbar contribution in both the SR and the W+jets CRs from the b-tagged CR.

The \Zvvjets~ background prediction in the SR is determined from the dimuon and dielectron CRs through transfer factors ($T^{\ell\ell}$).
They are obtained from simulation and account for the difference in the branching fractions of \Zvv~and $\Z\to \ell\ell$ decays and the impacts of lepton acceptance and selection efficiencies.
As the branching fraction of the \Z boson to electrons and muons is approximately a factor of three smaller than to neutrinos, the resulting constraint on the \Zvvjets~background from the dilepton CRs is limited by the statistical uncertainty in the dilepton control samples at large values of $\pt^\text{recoil}$.

The \phojets~CR is also used to constrain the \Zvvjets~background prediction via a transfer factor $T^\gamma$, which accounts for the difference in cross section and the acceptance and efficiency of identifying photon events.
This production mode is similar to that of \Zvvjets, providing thereby a constraint from data on the shape of the predicted Z \pt spectrum.
Since the production cross section for \phojets~events is roughly twice that for \Zvvjets~events, the addition of this CR to constrain the \Zvvjets~background prediction reduces the effect of the limited statistical power of the dilepton events.
However, additional theoretical systematic uncertainties are introduced in the extrapolation from this CR to the SR.

A further constraint on the \Zvvjets~background is given by W+jets events in the single lepton b-vetoed CRs via $T^{\PW/\Z}$ transfer factors.
Additional theoretical uncertainties are included for covering the extrapolation from $\Wlvjets$ to $\Zvvjets$ events.

\subsection{Systematic uncertainties}\label{sec:unc}

The $\pt^\text{recoil}$ spectra of the processes considered are determined through a binned maximum-likelihood fit, performed simultaneously across all fourteen CRs and two SRs.
Systematic uncertainties are treated as nuisance parameters $\boldsymbol{\theta}$ that are constrained in the fit.

Uncertainties associated with the transfer factors $T^{X}$ as a function of $\pt^\text{recoil}$ are each modeled with a Gaussian prior distribution.
They include theoretical uncertainties in the ratio of $\gamma$ and Z differential cross sections and in the ratio of W and Z differential cross sections, coming from the choice of the renormalization and factorization scales.
We also account for variations of $T^{X}$ due to the PDF uncertainties, following the NNPDF3.0 prescription~\cite{Ball:2014uwa}.
We consider uncertainties on $T^X$ associated with the electroweak corrections to $\gamma$, Z, and W processes, due to higher-order electroweak effects~\cite{Denner:2009gj,Denner:2011vu,Denner:2012ts,Kallweit:2015dum,Kuhn:2004em,Kuhn:2005gv,Kuhn:2005az,Kuhn:2007cv}.
Each of the uncertainties from renormalizaton and factorization scales, PDF, and electroweak effects is correlated among bins of \ptmiss, but is not correlated among different processes.
Finally, uncertainties in the efficiencies of b tagging AK4 jets and subjets are propagated as uncertainties on $T^{X}$.

The uncertainties detailed in the following only affect the normalizations of the respective processes and are given a log-normal prior distribution.

An uncertainty of 21\% in the heavy-flavor fraction in W+jets events is computed using CMS measurements of inclusive W+jets \cite{Khachatryan:2014uva} and W+heavy-flavor \cite{Khachatryan:2014uva,Chatrchyan:2013uza} production.
This is propagated to each of the SRs and the CRs by scaling up and down the heavy-flavor fractions in the prediction by one standard deviation.
These W+heavy-flavor uncertainties are correlated among all regions in the fit.
A similar method is used for the Z+heavy-flavor fraction  uncertainty (22\%) using measurements of Z+jets production at CMS \cite{Khachatryan:2014zya,Chatrchyan:2014dha}.
This uncertainty is also correlated among all regions, but is uncorrelated with the W+heavy-flavor uncertainty.
The magnitudes of these W$/$Z+heavy-flavor uncertainties are different for each region (depending on b tagging requirements) and range from 3 to 4\% of the nominal W$/$Z+jets prediction.

Additional uncertainties are included to account for the differences between data and simulation in the CA15 jet $m_\mathrm{SD}$ and BDT distributions.
To derive the uncertainty for top quark jets, the efficiency of the mass window and BDT selection is measured in data using the mass spectrum of CA15 jets observed in a CR that is enriched in \ttbar events, where one top quark decays to a muon and jets.
Then, variations due to the parton shower algorithm, higher-order corrections, and experimental effects are propagated to the efficiency measurement.
This is done for the loose and tight categories independently.
The final uncertainty for tagging CA15 jets from a top quark decay is found to be 6\% in both categories.
Similarly, the uncertainty in mistagging a q/g jet is measured by computing the efficiency in a Z$(\mu\mu)$+jets selection.
The mistag uncertainty is 7\%.
The CRs used to compute these efficiencies and uncertainties are those shown in Fig.~\ref{fig:toptag_data}.
The uncertainties corresponding to the $m_\mathrm{SD}$ and BDT distributions are only applied to the signal and minor-background predictions.
The same selection is applied in the SR and CRs for the data-driven backgrounds (Z+jets, W+jets, \ttbar), and so these uncertainties cancel in the transfer factors $T^X$.

Uncertainties in selection efficiencies amount to 1\% per selected muon, electron, or photon, and the uncertainty in the $\tau$ lepton veto is 3\%. These uncertainties are correlated across all $\pt^\text{recoil}$ bins.
A systematic uncertainty of 20\% is ascribed to the single top quark background prediction~\cite{Chatrchyan:1642680}, which is correlated among the SR and the leptonic CRs.
An uncertainty of 20\% is assigned to the diboson production cross section~\cite{Khachatryan:2016txa,Khachatryan:2016tgp}, and is correlated across all channels.

The QCD multijet background is estimated from MC simulation in all regions except for the $\gamma$+jets CR, where the prediction is obtained from a jet-to-photon misidentification probability measured in an independent control sample of events in data. An uncertainty of 100\% is used for the overall QCD multijet yield.
This uncertainty is estimated using a sample enriched in QCD multijet events, obtained by requiring the minimum azimuthal angle between $\ptvecmiss$ and the AK4 jet directions to be less than 0.1\,rad.

For processes estimated from MC simulation, \ptmiss uncertainties are obtained directly from simulation and propagated to $\pt^\text{recoil}$ following the standard CMS method~\cite{Khachatryan:2014gga}, which includes the application of uncertainties in jet energy corrections applied to AK4 jets and \ptmiss \cite{jec}.
The uncertainty in \ptmiss is used again as an uncertainty in the normalization in the final fit.

A systematic uncertainty of 2.5\%~\cite{CMS-PAS-LUM-17-001} in the integrated luminosity is included for all processes that are estimated using MC simulation.

The impact of statistical uncertainties on the predictions for simulation-driven backgrounds is negligible. For the transfer factors $T^X$, which are obtained from simulation and used to derive a data-driven estimate of the main backgrounds, we introduce additional nuisance parameters corresponding to bin-by-bin statistical uncertainties.

We further consider uncertainties in the signal cross sections, estimated by observing the effect of varying the renormalization and factorization scales by factors of 0.5 and 2.0, and of the PDF uncertainties.
To that end, an uncertainty of 10\% is assigned to the nonresonant signal cross sections.
The corresponding uncertainties in the resonant signal cross sections vary from 10\% to 32\% as a function of the mass of the scalar mediator.
Unlike other uncertainties, these are not propagated as nuisance parameters, but rather treated as uncertainties in the inclusive signal cross section.

\section{Results}

\begin{figure*}[!hbtp]\centering
      \includegraphics[width=0.49\textwidth]{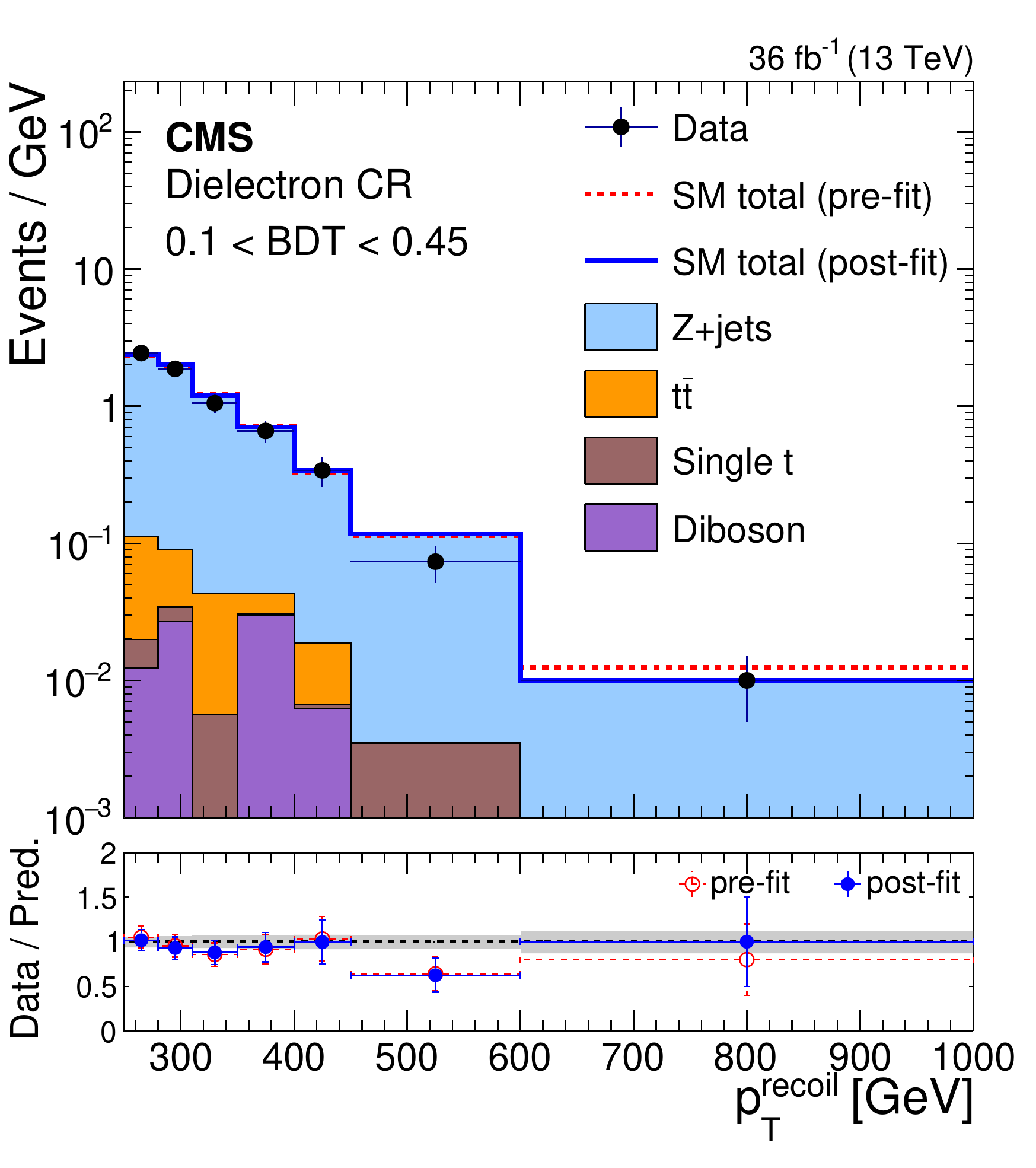}
      \includegraphics[width=0.49\textwidth]{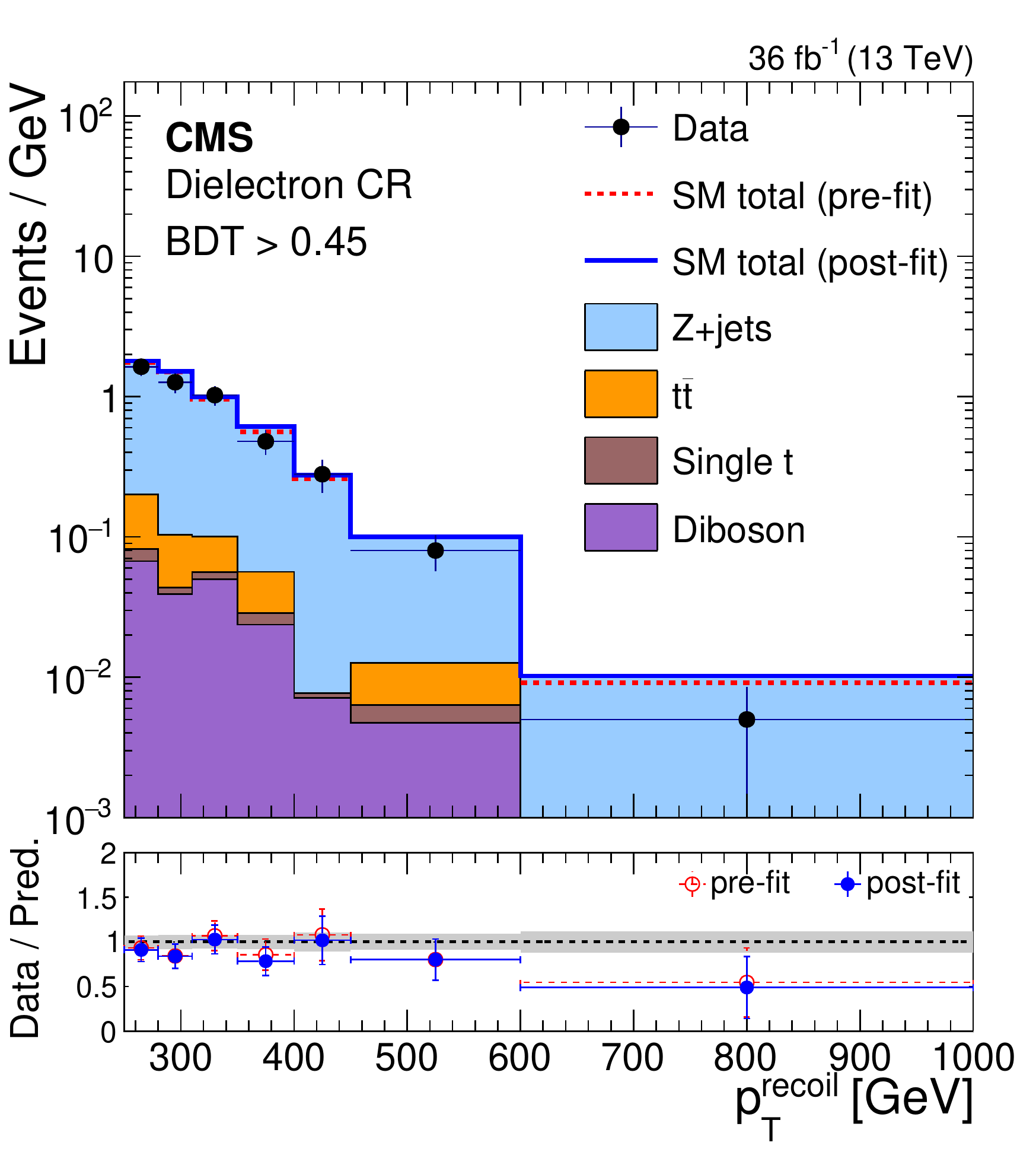}
      \includegraphics[width=0.49\textwidth]{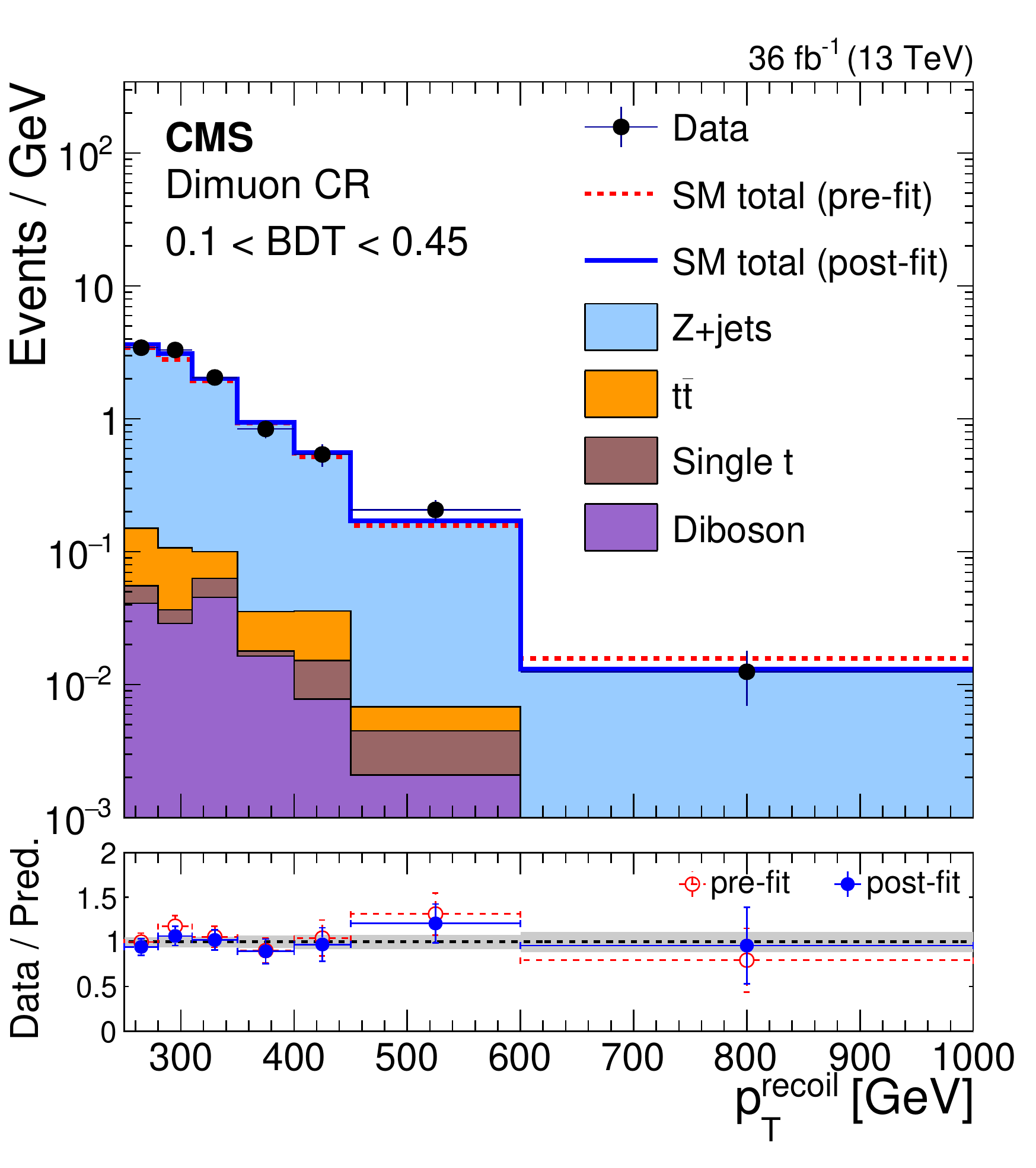}
      \includegraphics[width=0.49\textwidth]{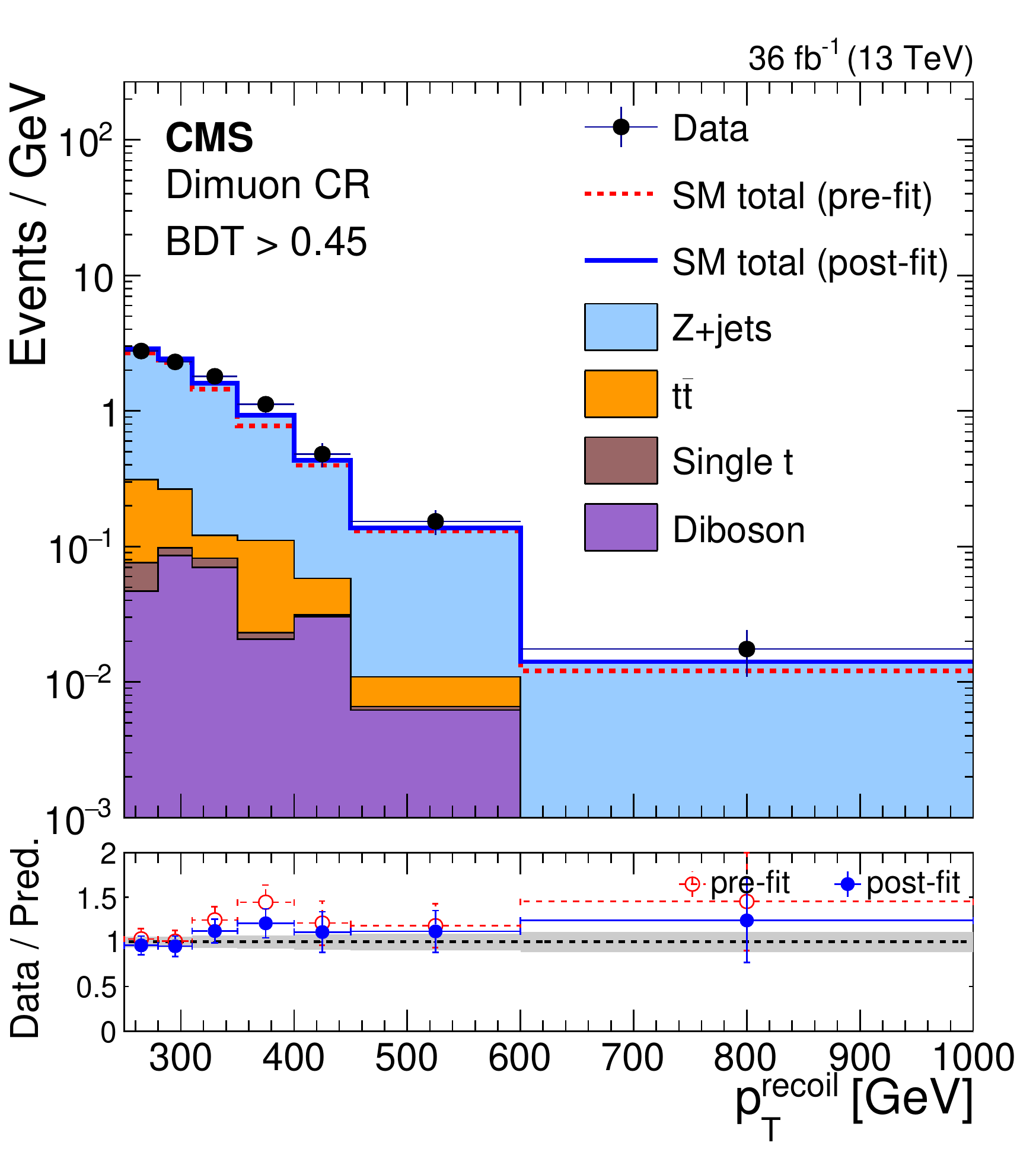}
    \caption{
      Comparison between data and SM predictions in the dilepton control regions before and after performing the simultaneous fit to the different control regions and signal region.
      Each bin shows the event yields divided by the width of the bin.
      The upper row of figures corresponds to the dielectron control region, and the lower row to the dimuon control region.
      The left (right) column of figures corresponds to the loose (tight) category of the control regions.
      The blue solid line represents the sum of the SM contributions normalized to their fitted yields. The red dashed line represents the sum of the SM contributions normalized to the prediction.
      The stacked histograms show the individual fitted SM contributions.
      The lower panel of each figure shows the ratio of data to fitted prediction.
      The gray band on the ratio indicates the one standard deviation uncertainty on the prediction after propagating all the systematic uncertainties and their correlations in the fit.
    }
 \label{fig:post_fit_plots_Zee}\end{figure*}

\begin{figure*}[!hbtp]\centering
      \includegraphics[width=0.49\textwidth]{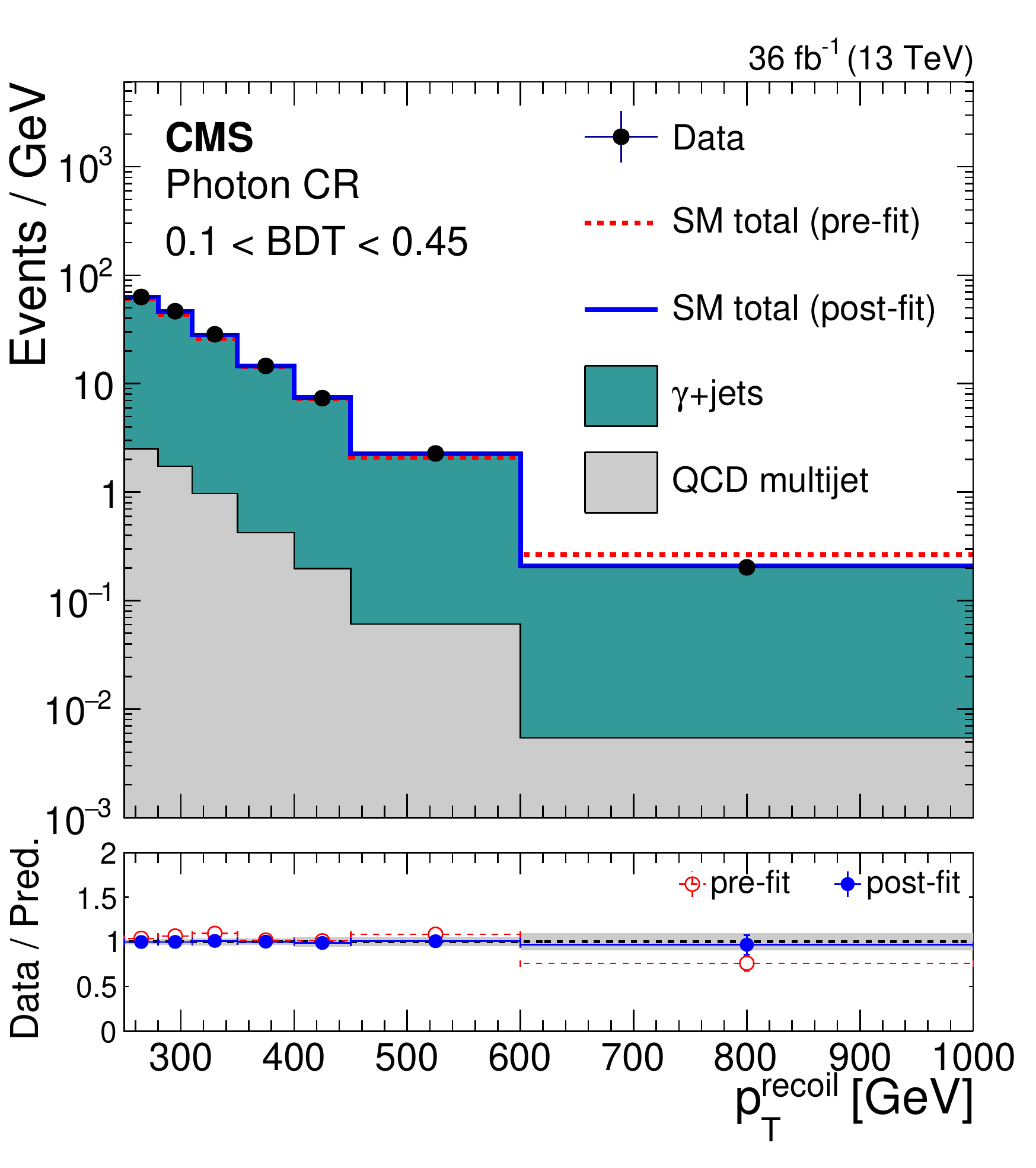}
      \includegraphics[width=0.49\textwidth]{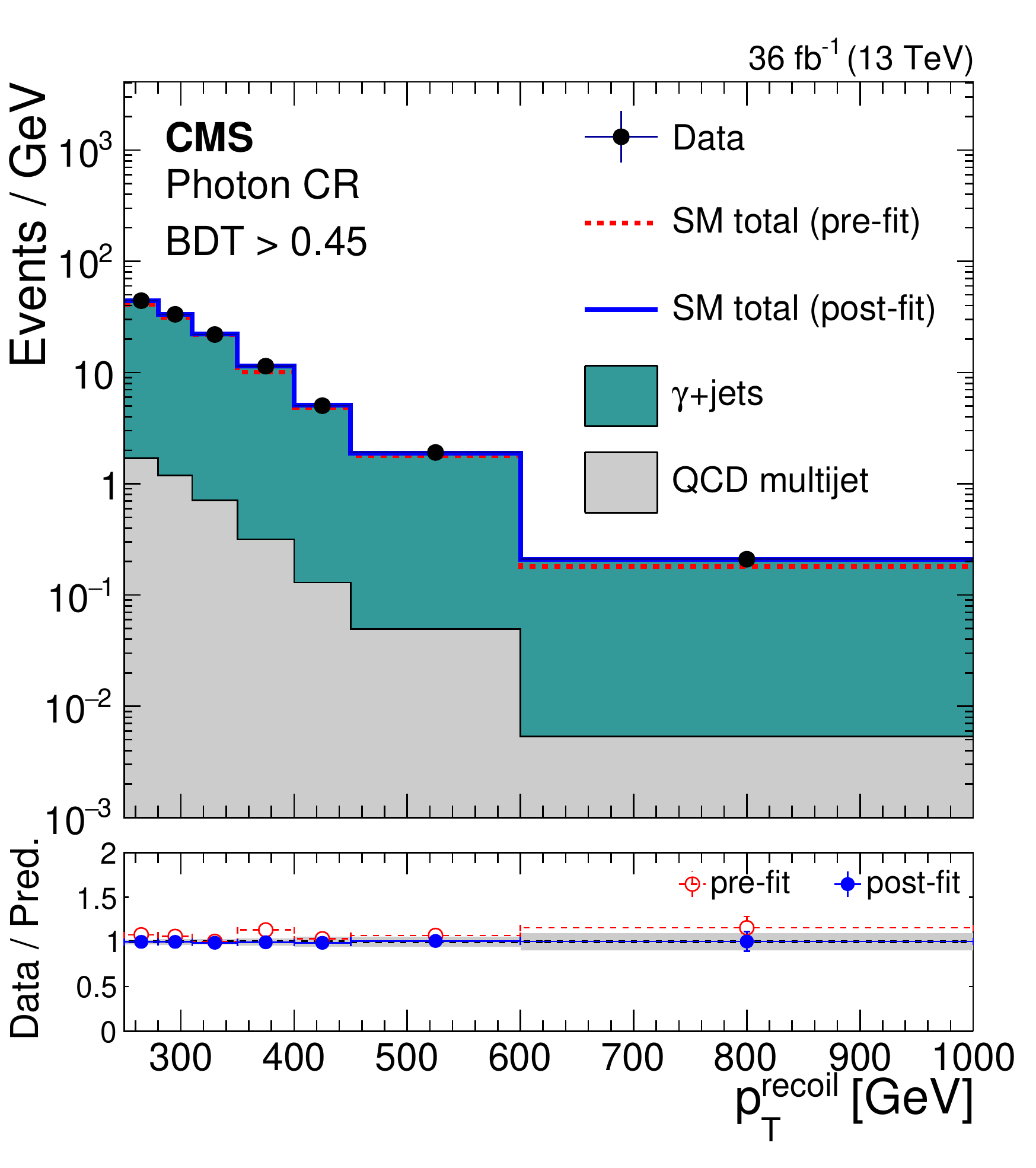}
    \caption{
      Comparison between data and SM predictions in the photon control regions before and after performing the simultaneous fit to the different control regions and signal region.
      Each bin shows the event yields divided by the width of the bin.
      The left (right) figure corresponds to the loose (tight) category of the control region.
      The blue solid line represents the sum of the SM contributions normalized to their fitted yields. The red dashed line represents the sum of the SM contributions normalized to the prediction.
      The stacked histograms show the individual fitted SM contributions.
      The lower panel of each figure shows the ratio of data to fitted prediction.
      The gray band on the ratio indicates the one standard deviation uncertainty on the prediction after propagating all the systematic uncertainties and their correlations in the fit.
    }
 \label{fig:post_fit_plots_Zmm}\end{figure*}

\begin{figure*}[!hbtp]\centering
      \includegraphics[width=0.49\textwidth]{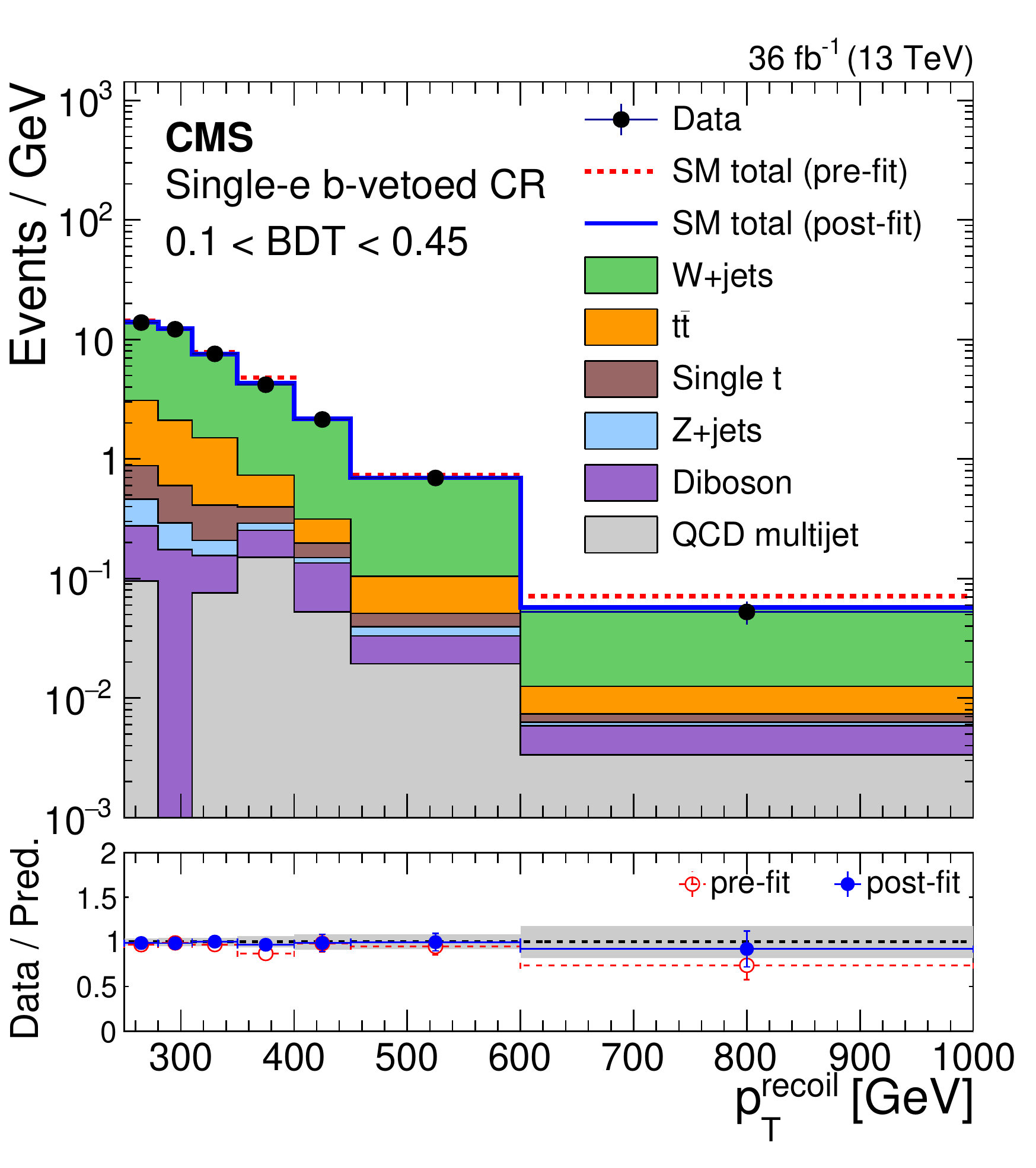}
      \includegraphics[width=0.49\textwidth]{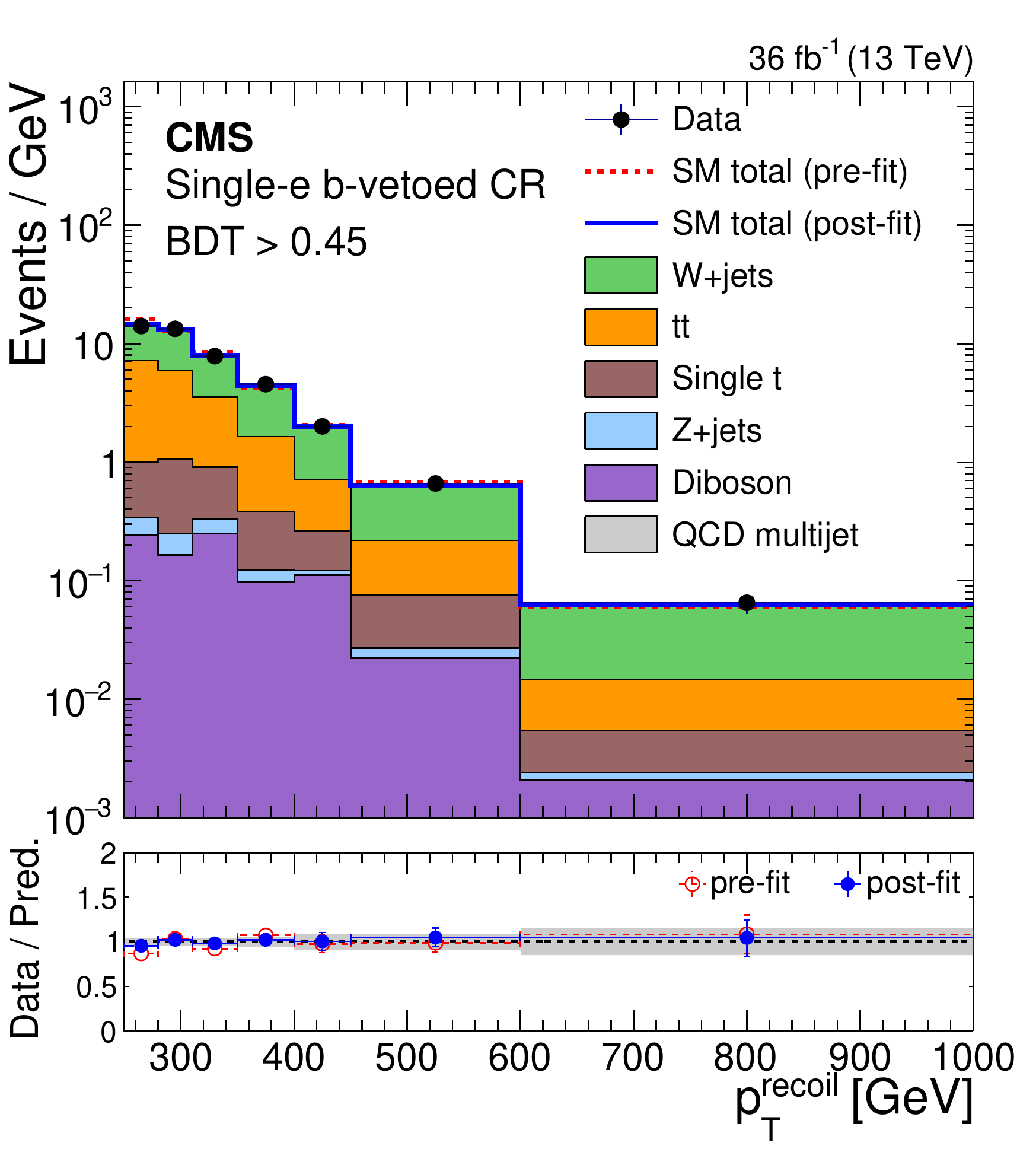}
    \\
      \includegraphics[width=0.49\textwidth]{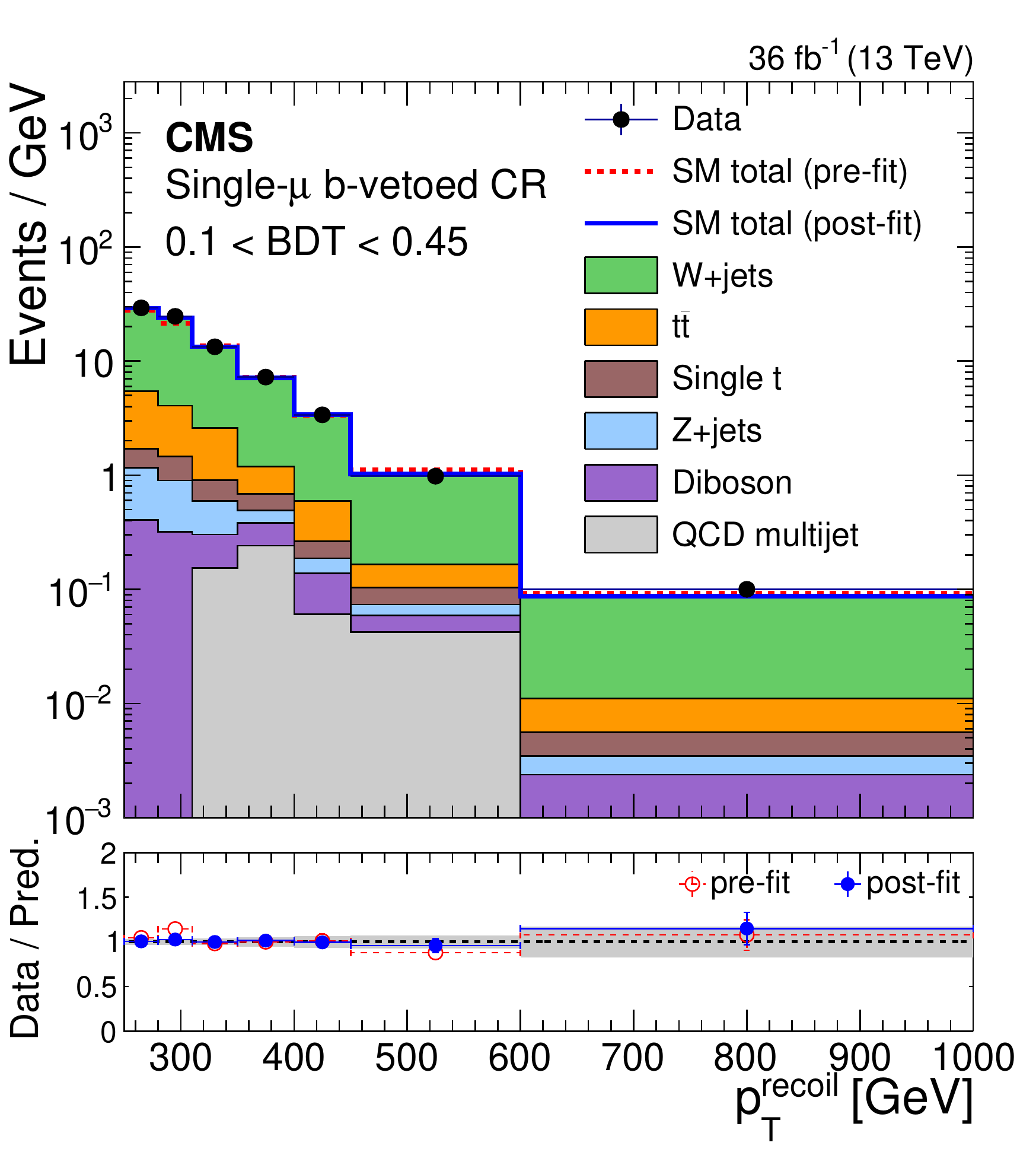}
      \includegraphics[width=0.49\textwidth]{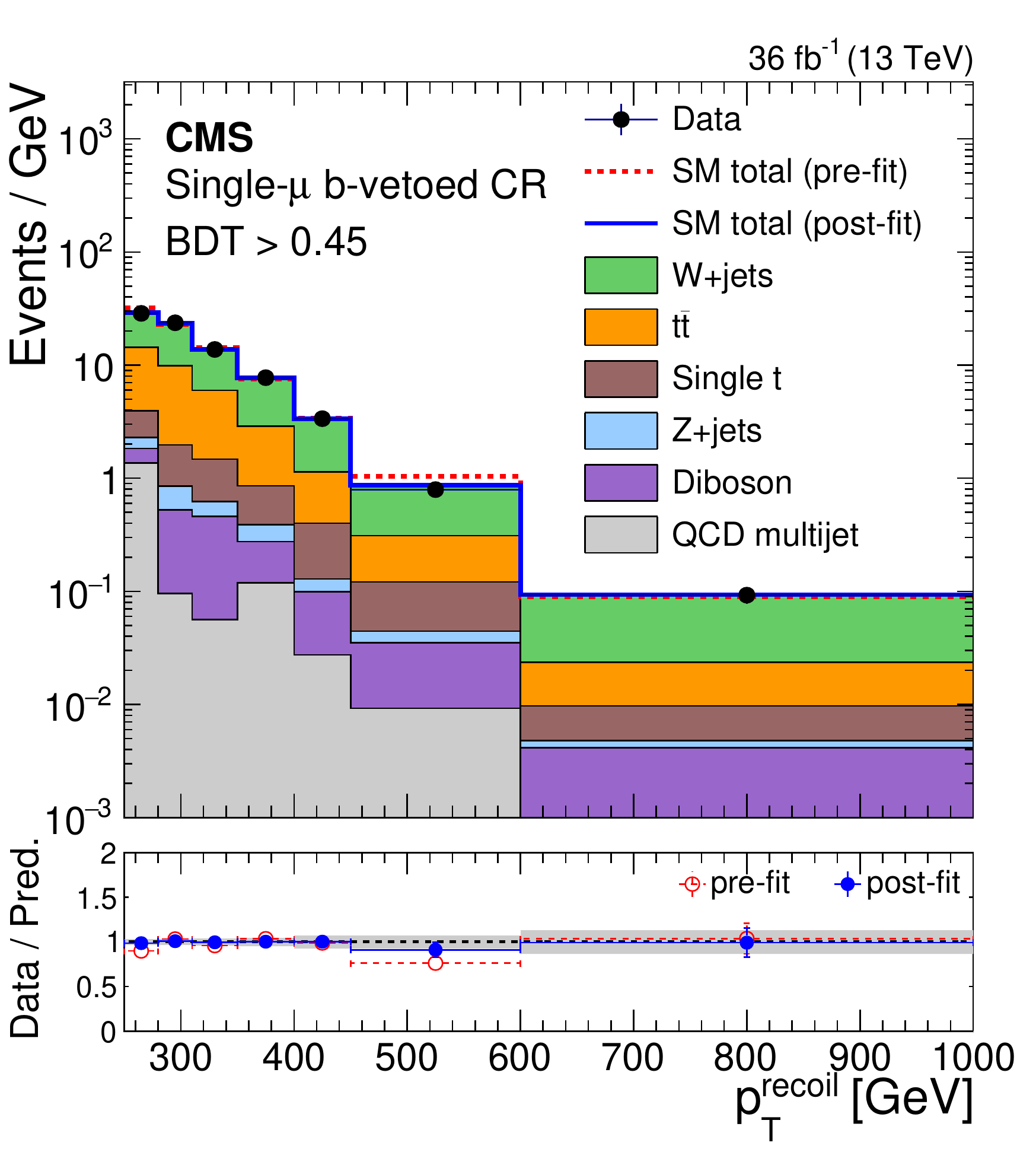}
    \\
    \caption{
      Comparison between data and SM predictions in the b-vetoed single lepton control regions before and after performing the simultaneous fit to the different control regions and signal region.
      Each bin shows the event yields divided by the width of the bin.
      The upper row of figures corresponds to the single electron b-vetoed control region, and lower row to the single muon b-vetoed control region.
      The left (right) column of figures corresponds to the loose (tight) category of the control regions.
      The blue solid line represents the sum of the SM contributions normalized to their fitted yields. The red dashed line represents the sum of the SM contributions normalized to the prediction.
      The stacked histograms show the individual fitted SM contributions.
      The lower panel of each figure shows the ratio of data to fitted prediction.
      The gray band on the ratio indicates the one standard deviation uncertainty on the prediction after propagating all the systematic uncertainties and their correlations in the fit.
    }
 \label{fig:post_fit_plots_W}\end{figure*}

\begin{figure*}[!hbtp]\centering
      \includegraphics[width=0.49\textwidth]{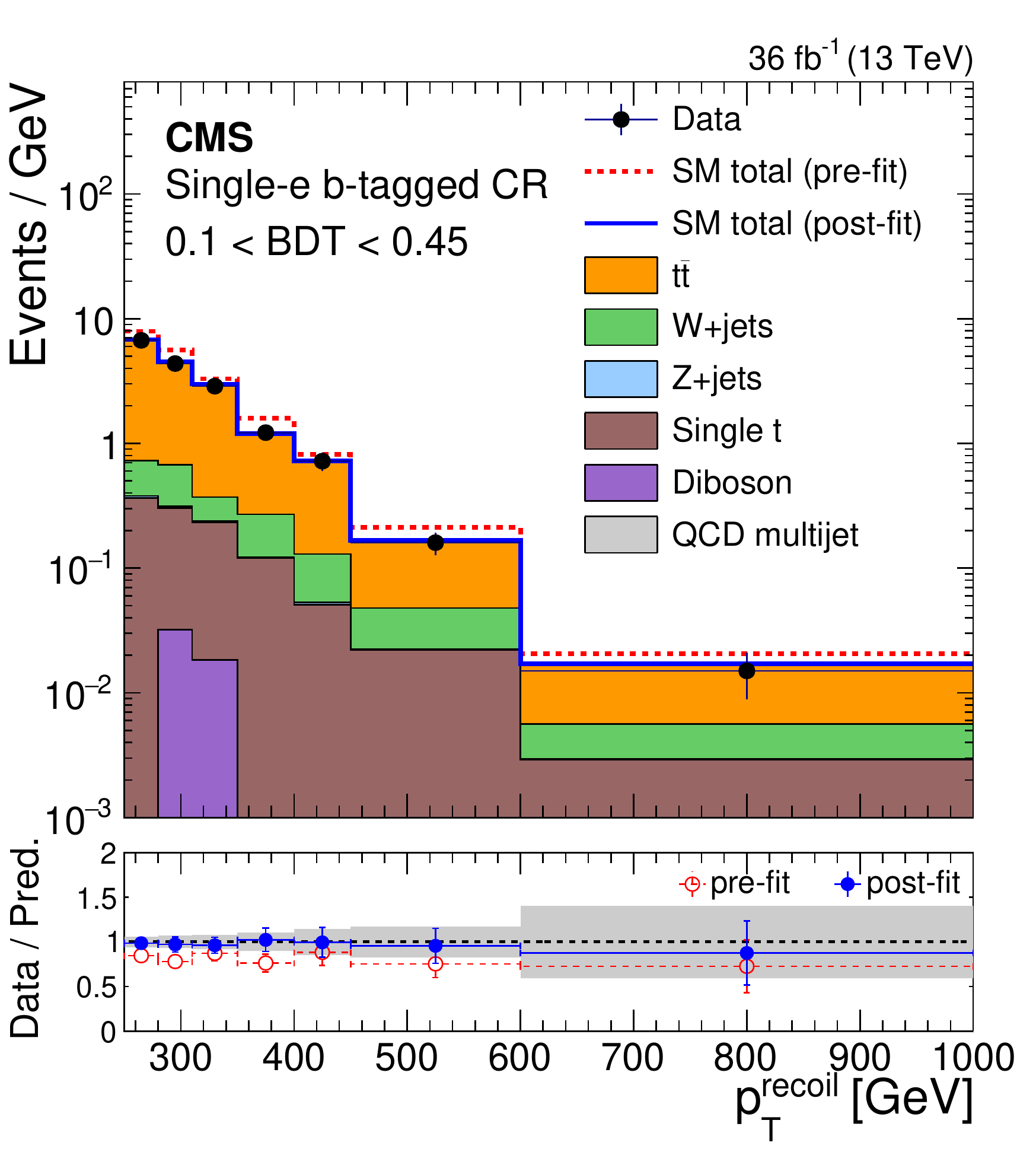}
      \includegraphics[width=0.49\textwidth]{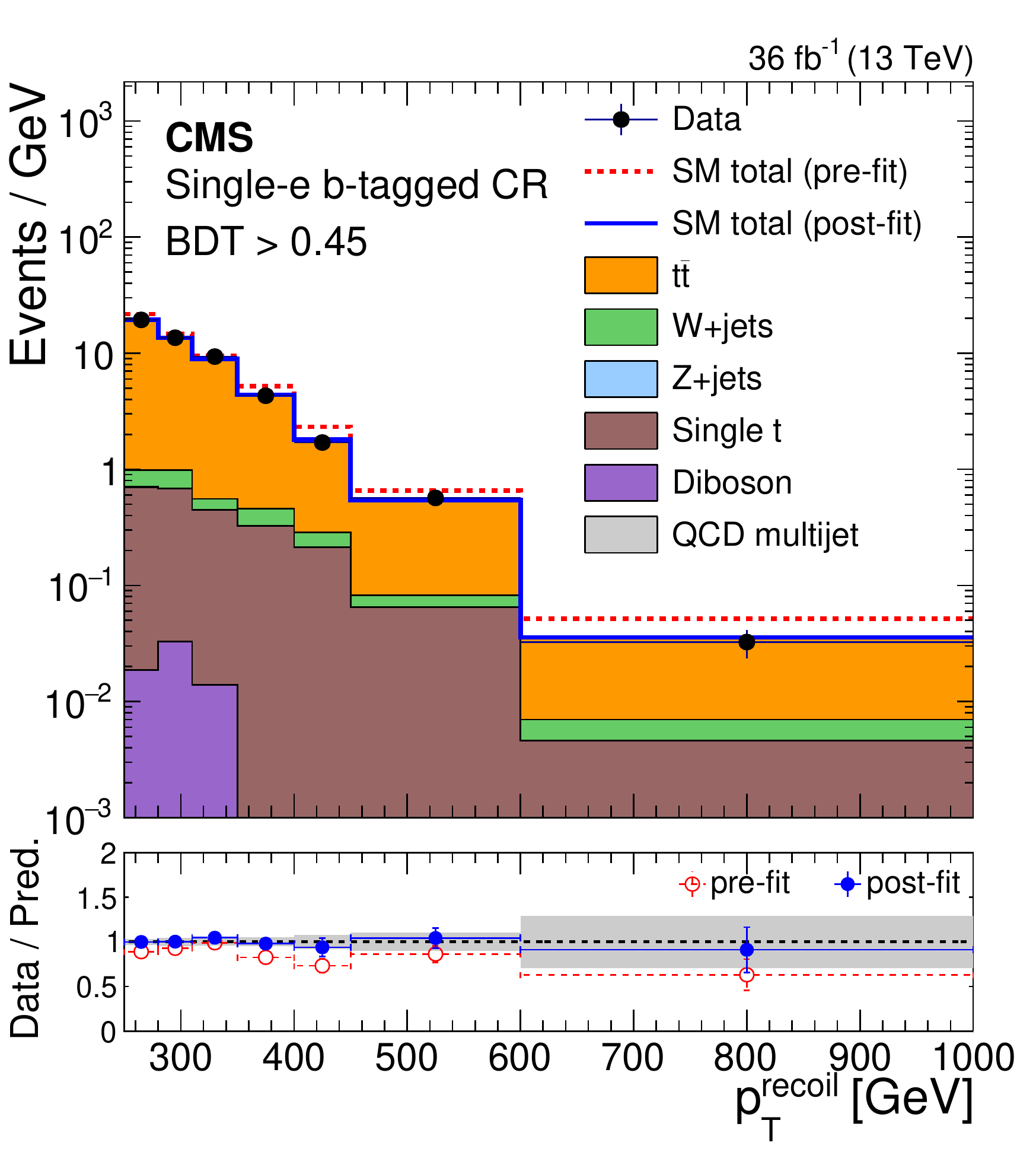}
    \\
      \includegraphics[width=0.49\textwidth]{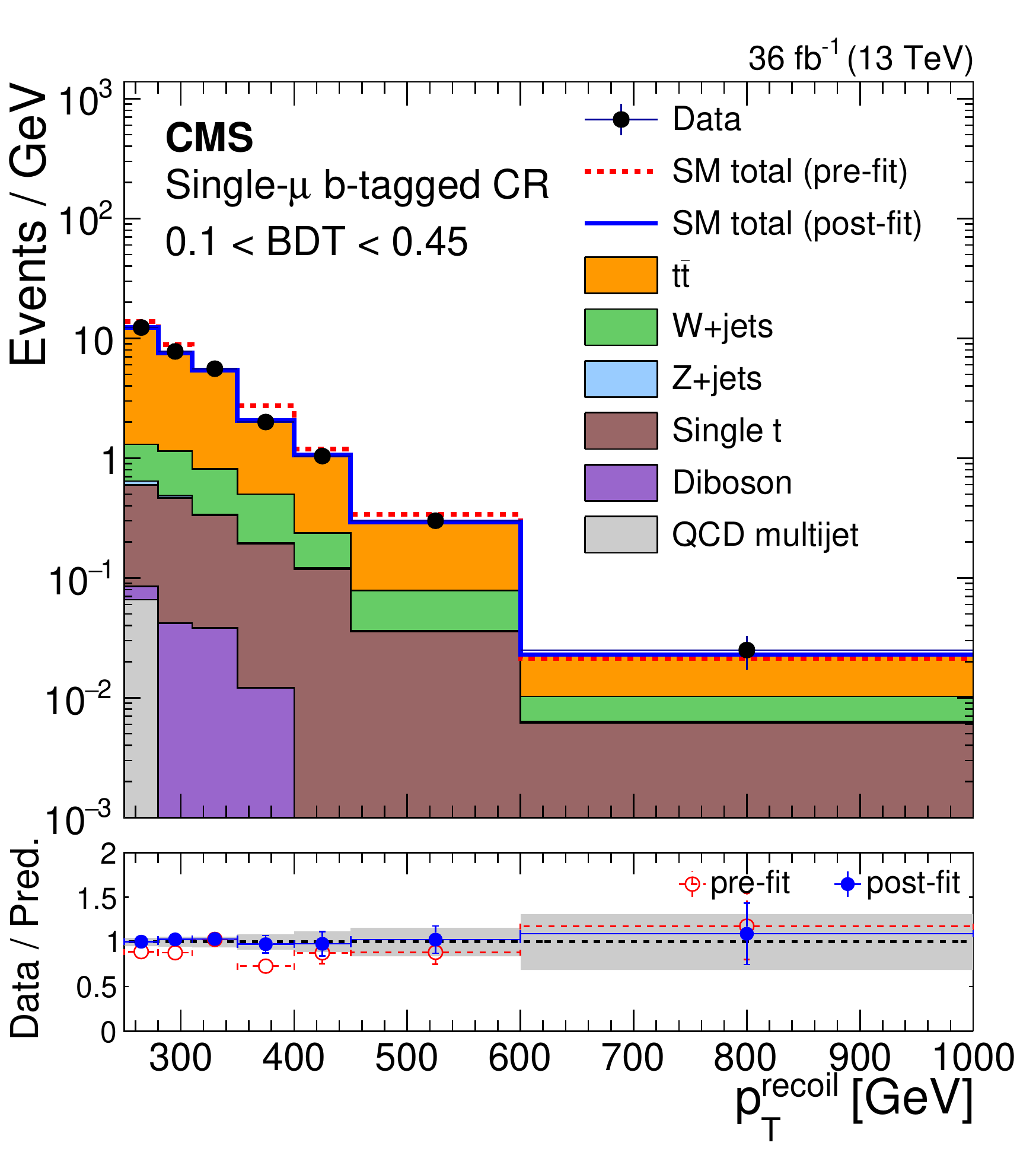}
      \includegraphics[width=0.49\textwidth]{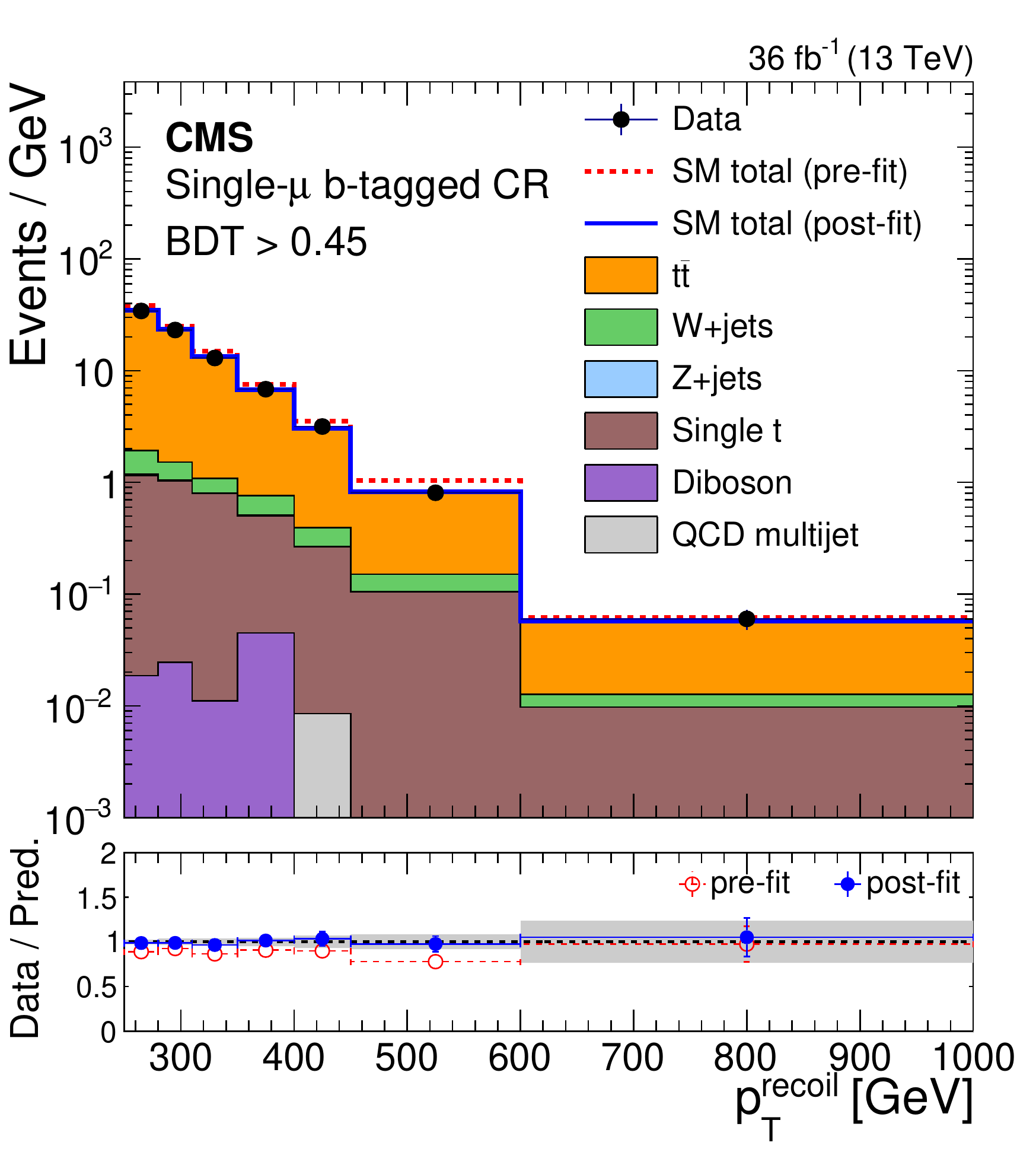}
    \\
    \caption{
      Comparison between data and SM predictions in the b-tagged single lepton control regions before and after performing the simultaneous fit to the different control regions and signal region.
      Each bin shows the event yields divided by the width of the bin.
      The upper row of figures corresponds to the single electron b-tagged control region, and lower row to the single muon b-tagged control region.
      The left (right) column of figures corresponds to the loose (tight) category of the control regions.
      The blue solid line represents the sum of the SM contributions normalized to their fitted yields. The red dashed line represents the sum of the SM contributions normalized to the prediction.
      The stacked histograms show the individual fitted SM contributions.
      The lower panel of each figure shows the ratio of data to fitted prediction.
      The gray band on the ratio indicates the one standard deviation uncertainty on the prediction after propagating all the systematic uncertainties and their correlations in the fit.
    }
 \label{fig:post_fit_plots_top}\end{figure*}

\begin{figure*}[!hbtp]\centering
    \includegraphics[width=0.49\textwidth]{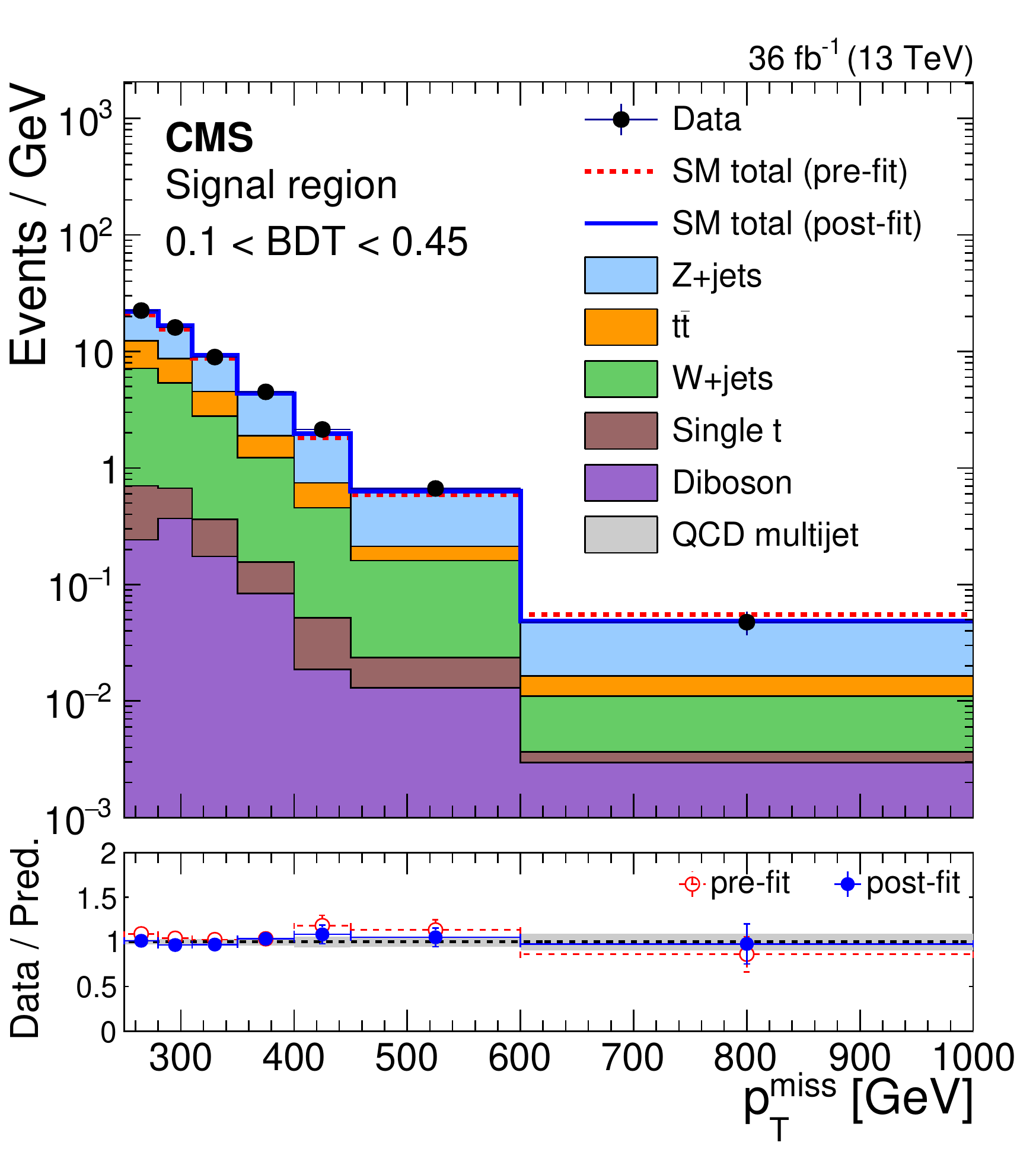}
    \includegraphics[width=0.49\textwidth]{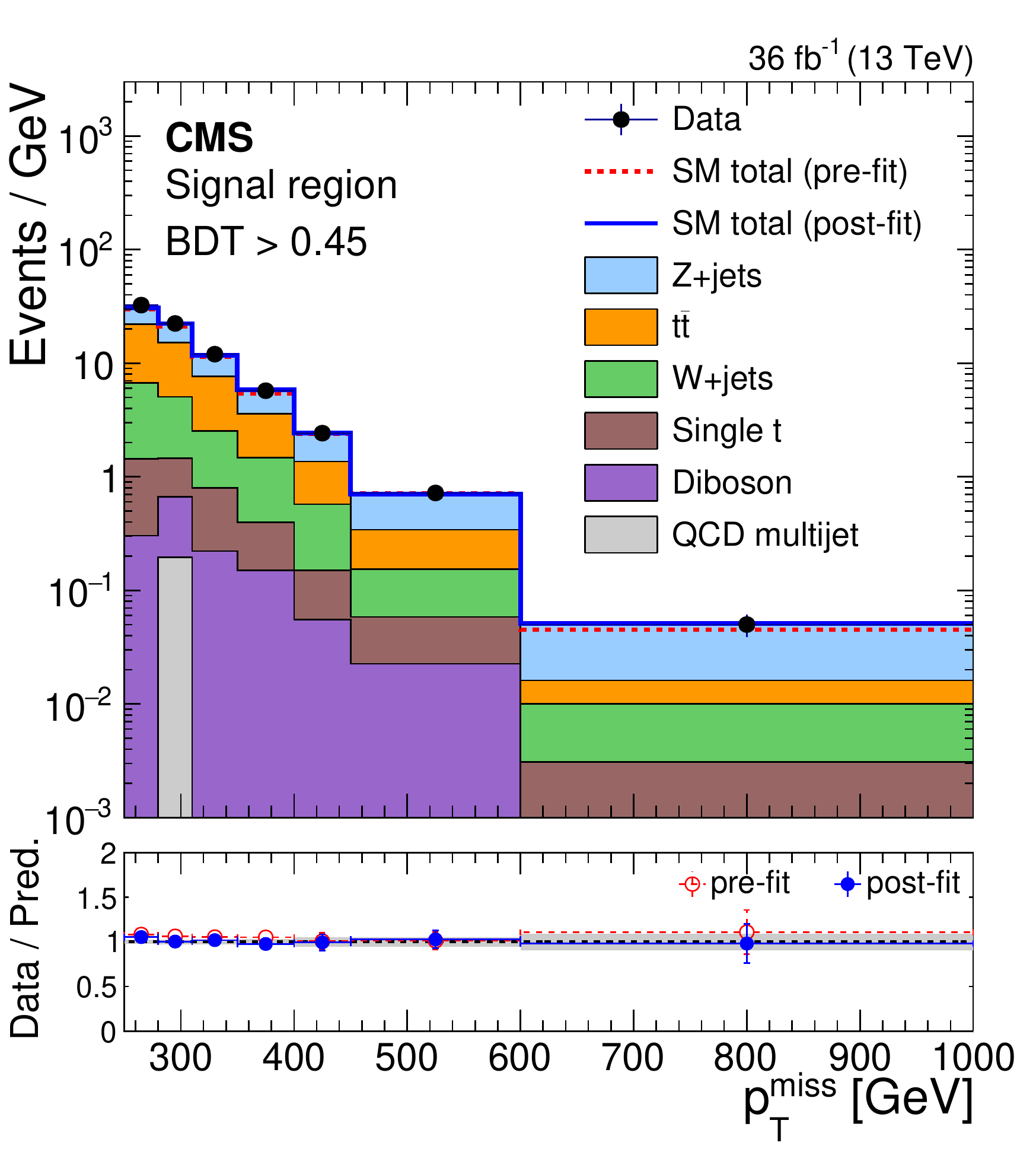}
  \caption{
    Distribution of \ptmiss from SM backgrounds and data in the signal region after simultaneously fitting the signal region and all control regions.
    Each bin shows the event yields divided by the width of the bin.
    The left (right) figure corresponds to the loose (tight) category of the signal region.
    The stacked histograms show the individual fitted SM background contributions. The blue solid line represents the sum of the SM background contributions normalized to their fitted yields. The red dashed line represents the sum of the SM background contributions normalized to the prediction.
      The lower panel of each figure shows the ratio of data to fitted prediction.
      The gray band on the ratio indicates the one standard deviation uncertainty on the prediction after propagating all the systematic uncertainties and their correlations in the fit.
  }
 \label{fig:post_fit_plots}\end{figure*}

Figures~\ref{fig:post_fit_plots_Zee}--\ref{fig:post_fit_plots_top} show the results of the simultaneous fit in all fourteen control regions and two signal regions.
The distributions observed in all CRs agree with predictions.
Figure~\ref{fig:post_fit_plots} shows the distribution in $\ptmiss$ in the signal region  under the background-only hypothesis.
Data are found to be in agreement with the SM prediction.
The fit does not require any nuisance parameter to vary more than 1.2 standard deviations from its initial value.

\begin{figure}[htbp]
  \centering
  \includegraphics[width=0.75\textwidth]{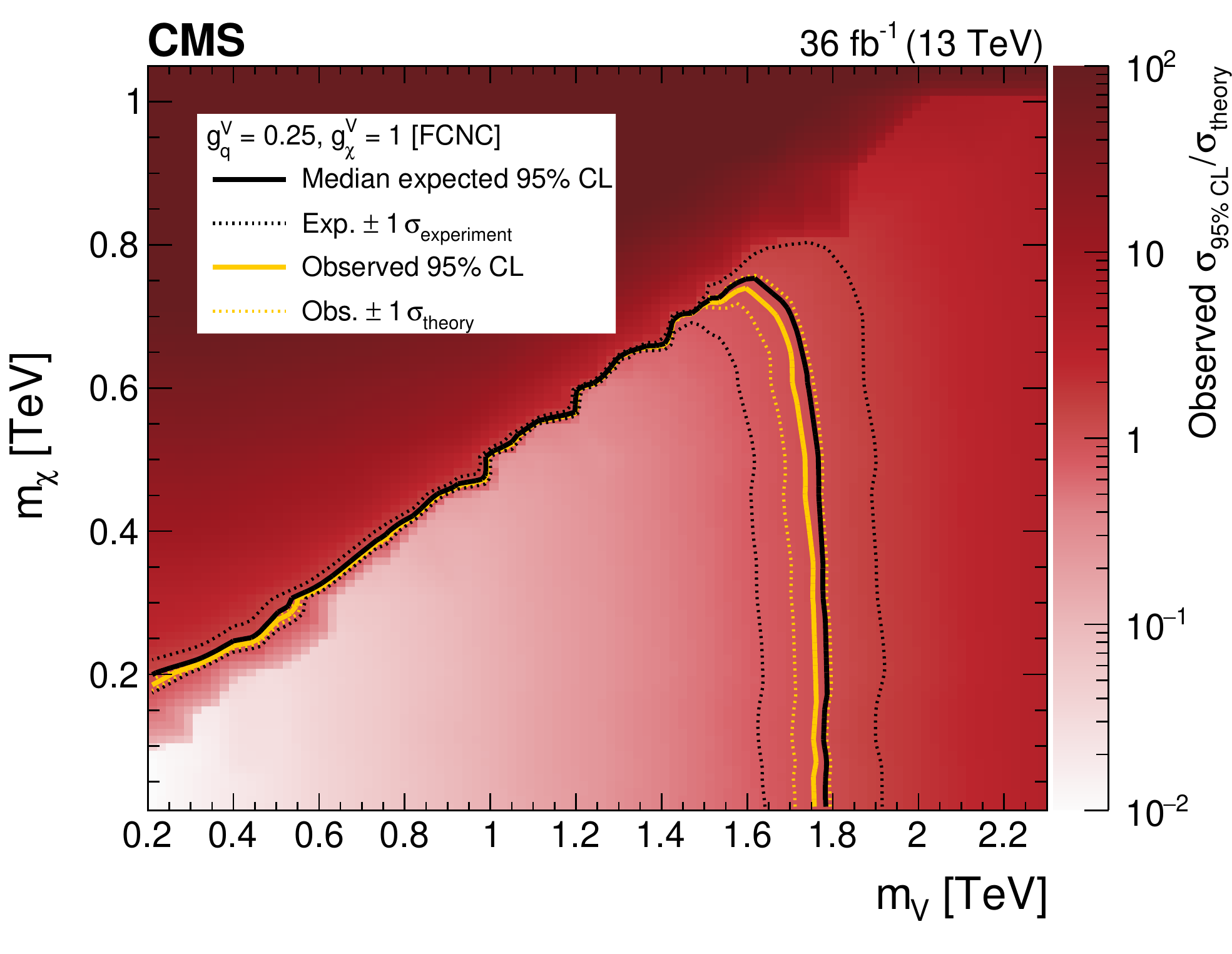}
  \caption{Results for the FCNC interpretation presented in the two-dimensional plane spanned by the mediator and DM masses.
  The mediator is assumed to have purely vector couplings to quarks and DM particles.
  The observed exclusion range (gold solid line) is shown. The gold dashed lines show the cases in which the predicted cross section is shifted by the assigned theoretical uncertainty.
  The expected exclusion range is indicated by a black solid line, demonstrating the search sensitivity of the analysis. The experimental uncertainties are shown in black dashed lines.}
  \label{fig:vlimits}
\end{figure}

\begin{figure}[htbp]
  \centering
  \includegraphics[width=0.75\textwidth]{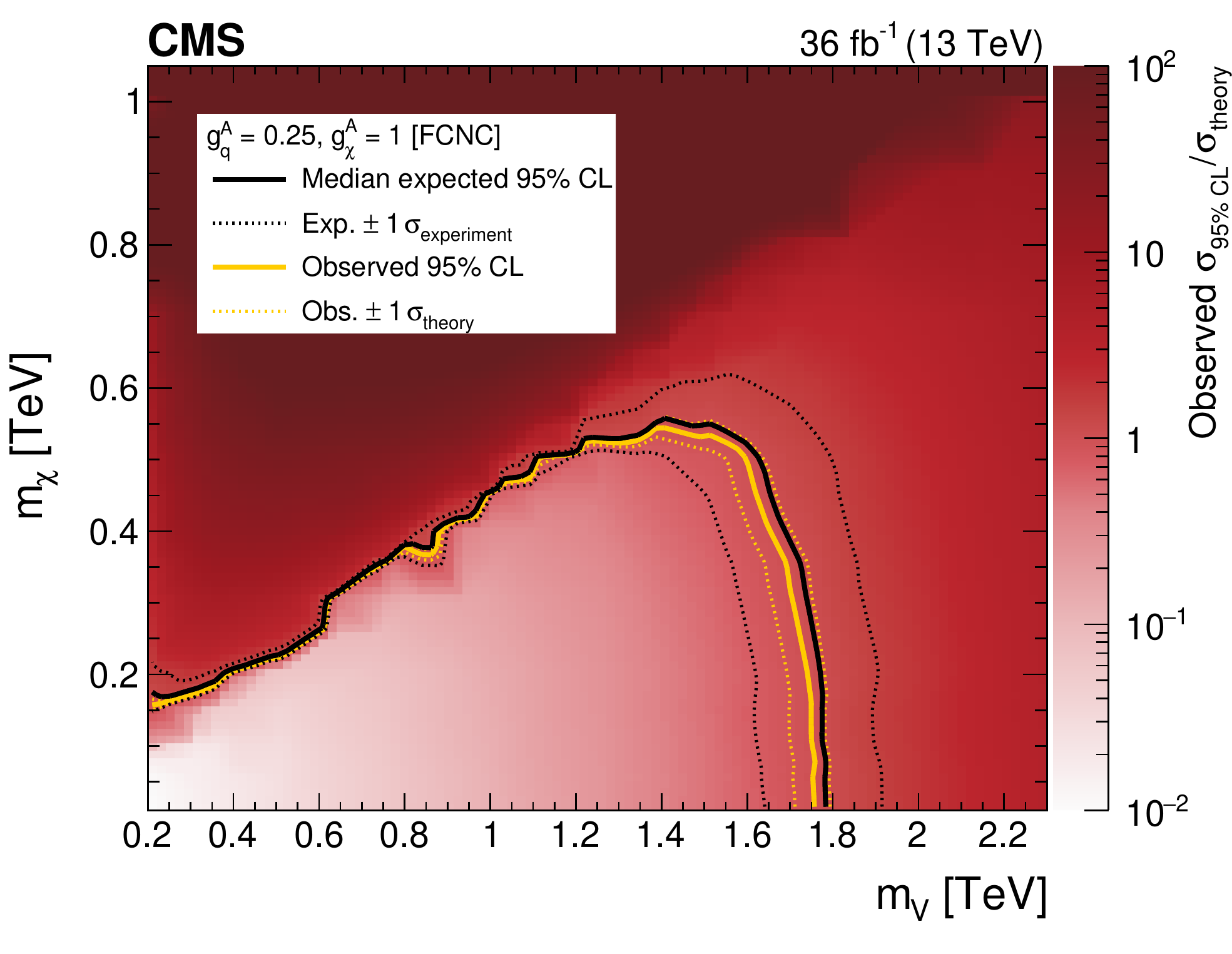}
  \caption{Results for the FCNC interpretation presented in the two-dimensional plane spanned by the mediator and DM masses.
  The mediator is assumed to have purely axial couplings to quarks and DM particles.
  The observed exclusion range (gold solid line) is shown. The gold dashed lines show the cases in which the predicted cross section is shifted by the assigned theoretical uncertainty.
  The expected exclusion range is indicated by a black solid line, demonstrating the search sensitivity of the analysis. The experimental uncertainties are shown in black dashed lines.}
  \label{fig:alimits}
\end{figure}

The results of the search are first interpreted in terms of the simplified model for monotop production via an FCNC.
Expected and observed limits at 95\% confidence level (CL) are set using the asymptotic approximation~\cite{asymptotic:cowan} of the CL$_\mathrm{s}$ criterion~\cite{cls:junk,Read:2002hq} with a profile likelihood ratio as the test statistic, in which systematic uncertainties are modeled as nuisance parameters.
Figure~\ref{fig:vlimits} shows the exclusion as a function of the mediator mass $m_\Vrm$ and DM particle mass $m_\chi$, assuming $g^\Vrm_\mathrm{q} = 0.25$, $g^\Vrm_\chi = 1$, and $g^\Arm_\mathrm{q} = g^\Arm_\chi = 0$.
At $m_\chi<100\GeV$, we observe that the result is roughly independent of $m_\chi$, and the range $0.2<m_\Vrm<1.75\TeV$ is excluded.
This can be compared to an expected exclusion of $0.2<m_\Vrm<1.78\TeV$.
At very high $m_\chi$ (\ie, $2m_\chi \gg m_\Vrm$), the parameter space is not excluded because the available phase space for the decay to DM decreases in this region.
Figure~\ref{fig:alimits} shows an analogous result, obtained with the assumptions $g^\Arm_\mathrm{q} = 0.25$, $g^\Arm_\chi = 1$, and $g^\Vrm_\mathrm{q} = g^\Vrm_\chi = 0$.
At $m_\chi\sim1\GeV$, the result in the axial case is very similar to the vector case.
An exclusion of $0.2<m_\Vrm<1.75\TeV$ is obtained for the FCNCs, compared to an expected exclusion of $0.2<m_\Vrm<1.78\TeV$.
However, as $m_\chi$ approaches the off-shell region, the shape of the exclusion is modified owing to the coupling structure.

\begin{figure}[htbp]
  \centering
  \includegraphics[width=0.75\textwidth]{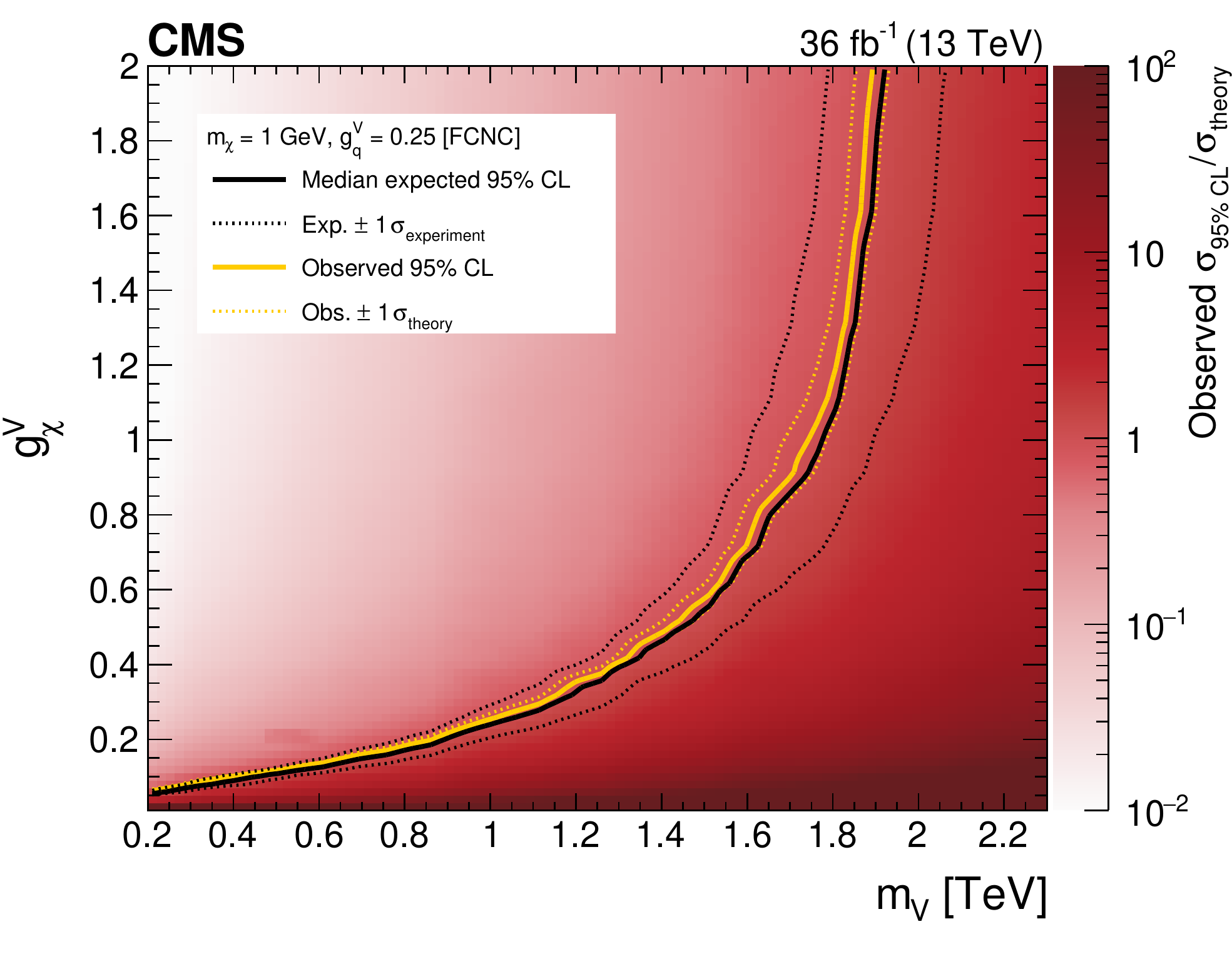} \\
  \includegraphics[width=0.75\textwidth]{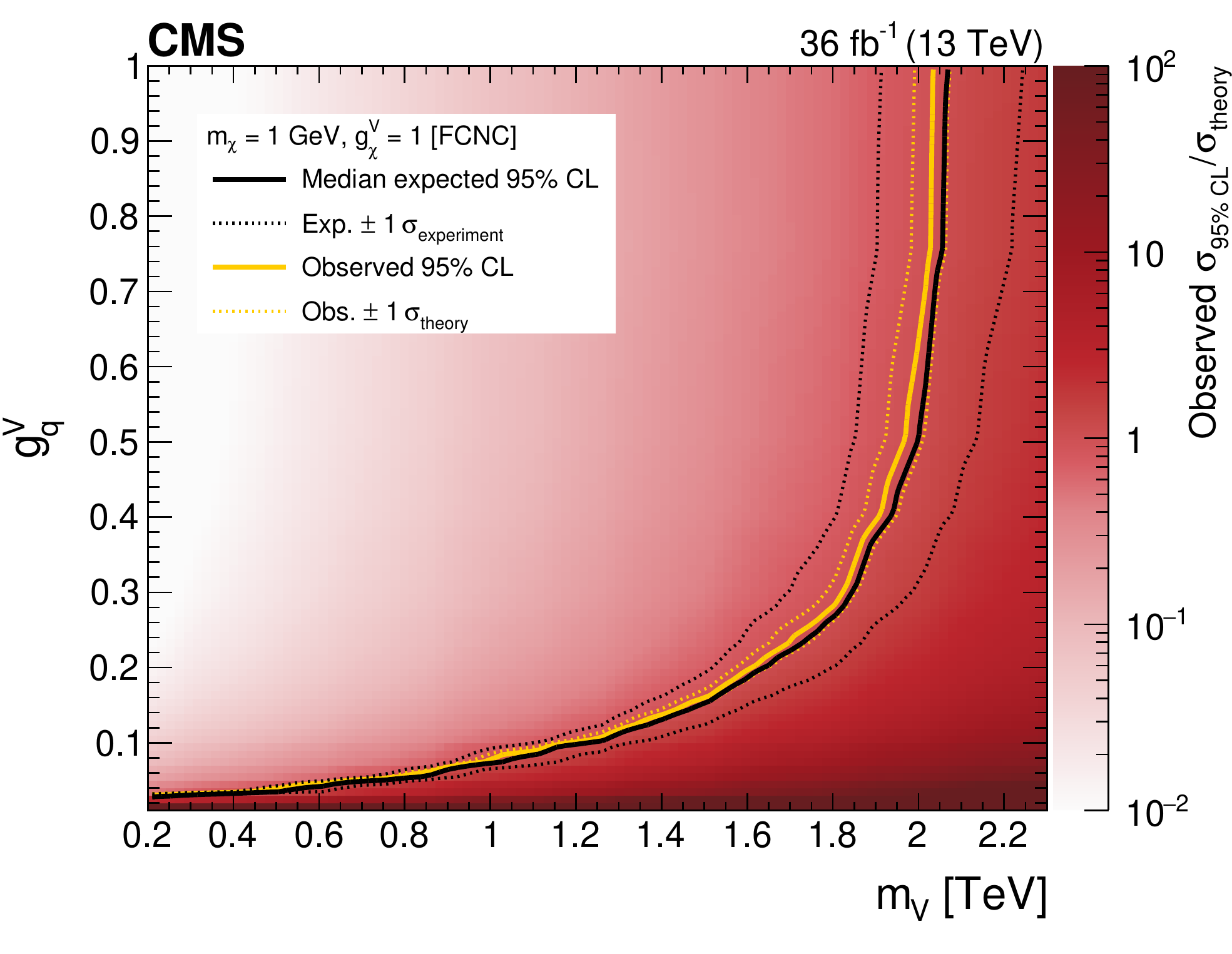}
  \caption{Results for the FCNC interpretation presented in the two-dimensional plane spanned by the mediator mass and the coupling between the mediator and DM (upper) or quarks (lower).
  The mediator is assumed to have purely vector couplings.
  The observed exclusion range (gold solid line) is shown. The gold dashed lines show the cases in which the predicted cross section is shifted by the assigned theoretical uncertainty.
  The expected exclusion range is indicated by a black solid line, demonstrating the search sensitivity of the analysis. The experimental uncertainties are shown in black dashed lines.}
  \label{fig:gvmlimits}
\end{figure}

\begin{figure}[htbp]
  \centering
  \includegraphics[width=0.75\textwidth]{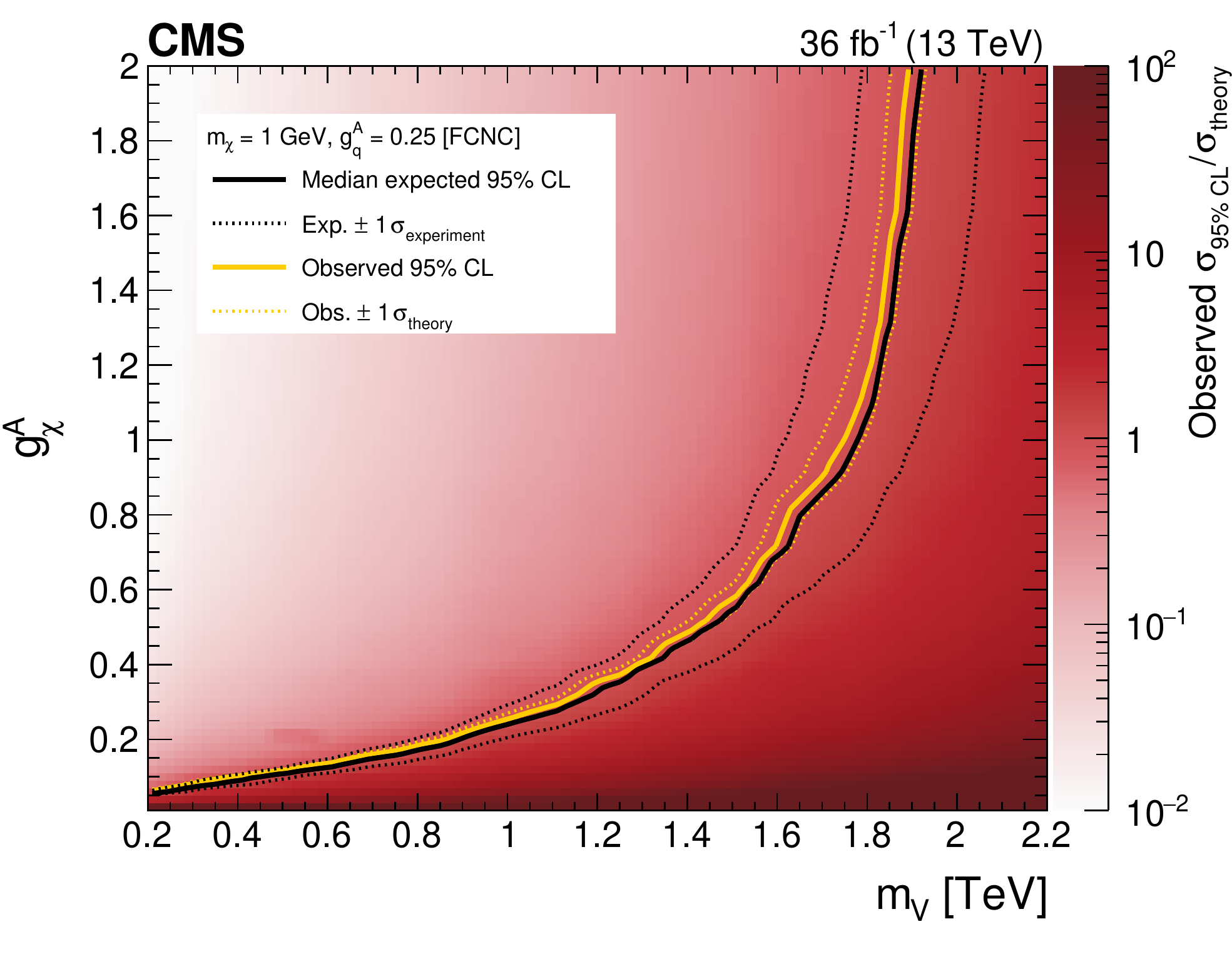} \\
  \includegraphics[width=0.75\textwidth]{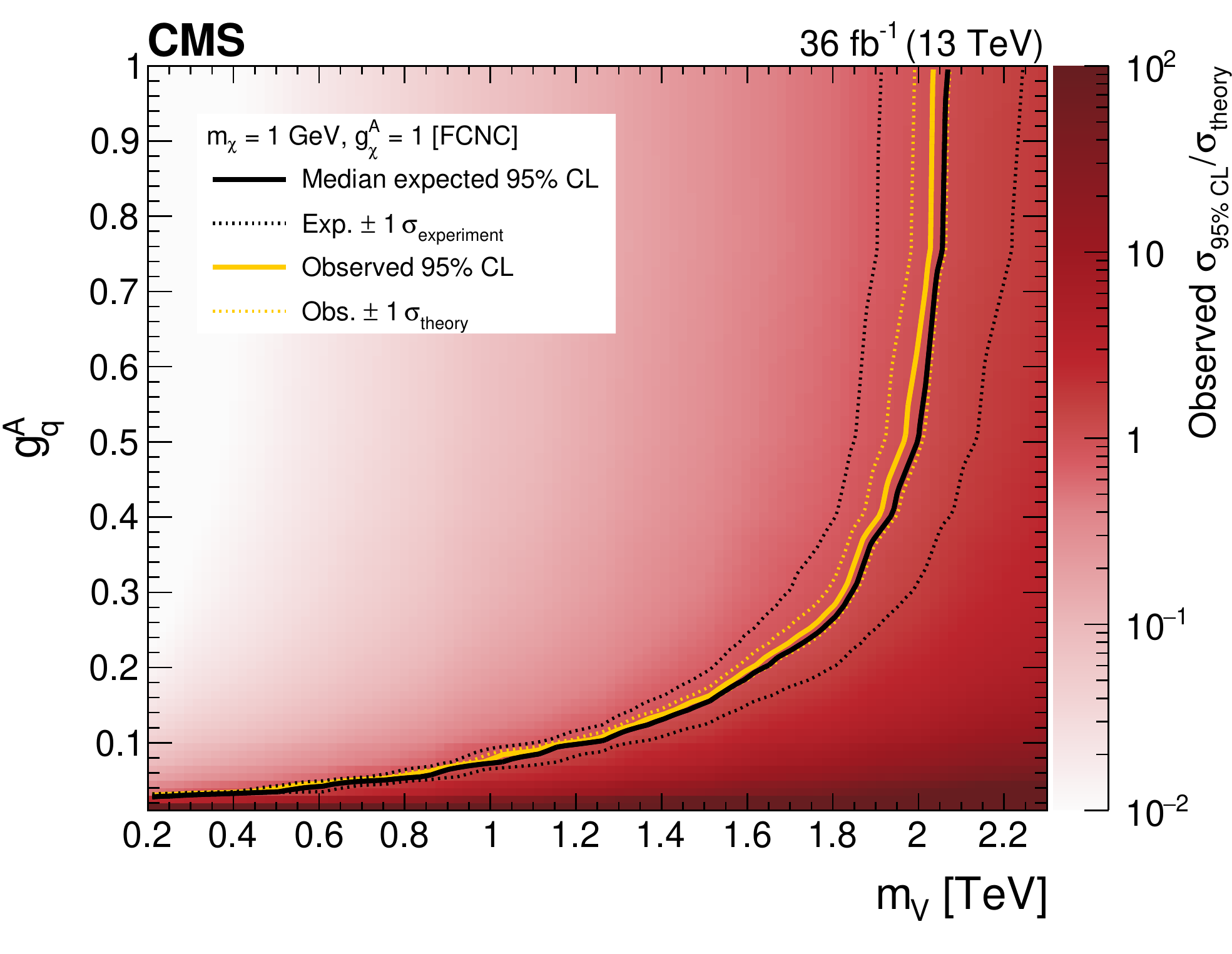}
  \caption{Results for the FCNC interpretation presented in the two-dimensional plane spanned by the mediator mass and the coupling between the mediator and DM (upper) or quarks (lower).
  The mediator is assumed to have purely axial couplings.
  The observed exclusion range (gold solid line) is shown. The gold dashed lines show the cases in which the predicted cross section is shifted by the assigned theoretical uncertainty.
  The expected exclusion range is indicated by a black solid line, demonstrating the search sensitivity of the analysis. The experimental uncertainties are shown in black dashed lines.}
  \label{fig:gamlimits}
\end{figure}

In addition to considering the dependence on the DM and mediator masses, limits are calculated as a function of the couplings between DM and the mediator, and between quarks and the mediator.
We fix $m_\chi=1\GeV$ and show the 95\% CL exclusion in the planes spanned by the couplings and $m_\Vrm$, assuming vector- (Fig.~\ref{fig:gvmlimits}) and axial-only couplings (Fig.~\ref{fig:gamlimits}).
Very little difference is observed between the two coupling schemes.
At low mediator and DM masses, coupling combinations as small as $g^{\Vrm,\Arm}_\chi=0.05, g^{\Vrm,\Arm}_\mathrm{q}=0.25$ and $g^{\Vrm,\Arm}_\chi=1,g^{\Vrm,\Arm}_\mathrm{q}=0.05$ are excluded.
Fig.~\ifthenelse{\boolean{cms@external}}{7}{\ref{fig:3d}} in \suppMaterialii{} shows the maximum excluded $m_\Vrm$ as a function of $g^\Vrm_\chi$ and $g^\Vrm_\mathrm{q}$.

\begin{figure}[htbp]
  \centering
  \includegraphics[width=0.75\textwidth]{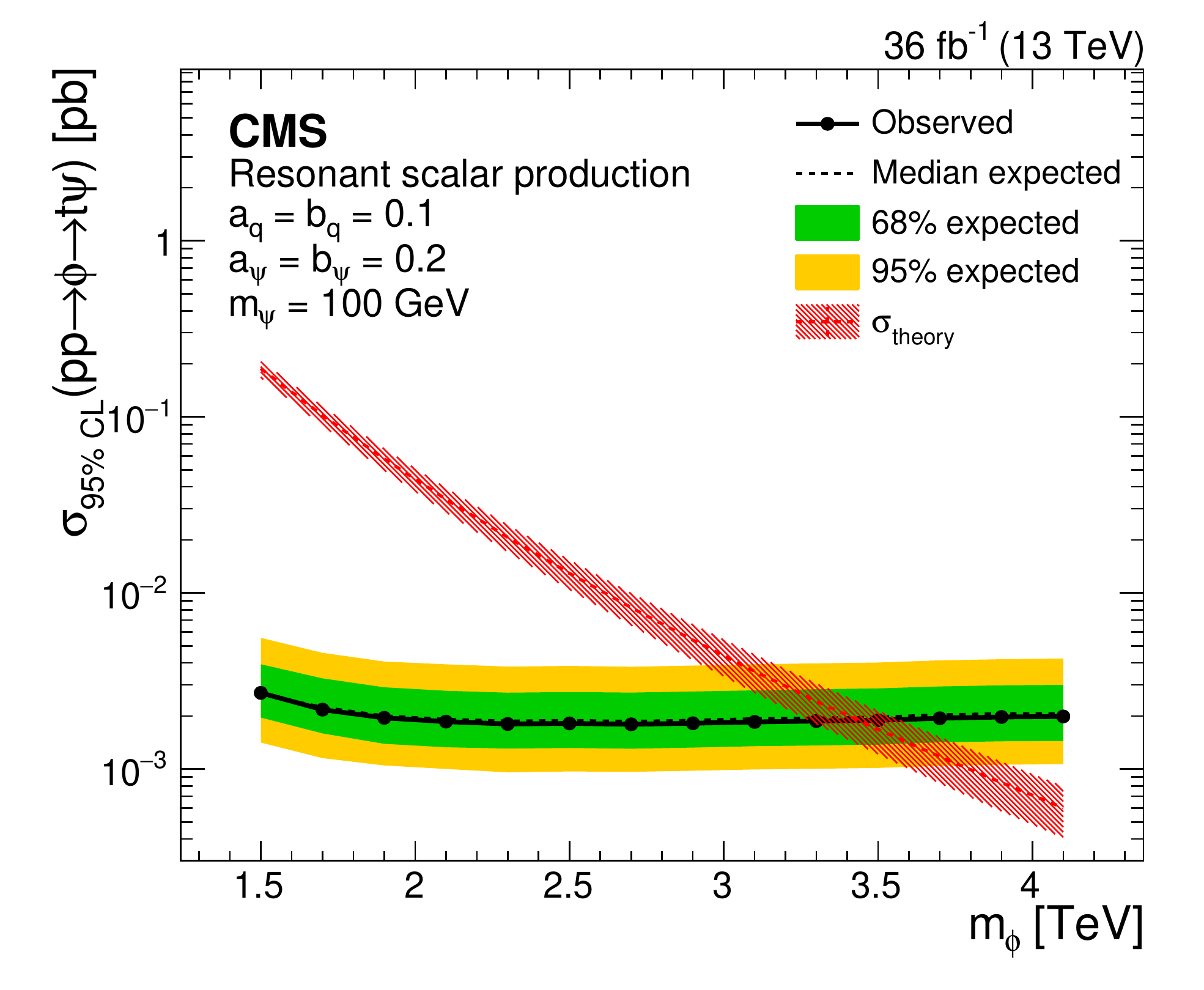}
  \caption{Upper limits at 95\% CL on the mass of the scalar particle $\phi$ in the resonant model, assuming fixed  $a_\mathrm{q} = b_\mathrm{q} = 0.1$ and $a_\psi = b_\psi=0.2$.
  The green and yellow bands represent one and two standard deviations of experimental uncertainties, respectively.
  The red hatched band represents the signal cross section uncertainty as a function of $m_\phi$.}
  \label{fig:res1D}
\end{figure}

Figure~\ref{fig:res1D} shows the results in the resonant model interpretation.
The DM mass is fixed at $m_\psi=100\GeV$, and the couplings are assumed to be $a_\mathrm{q} = b_\mathrm{q} = 0.1$ and $a_\psi = b_\psi = 0.2$.
Scalars with mass $1.5 < m_\phi<3.4\TeV$ are excluded at 95\% CL.

A summary of the importance of the systematic uncertainties is presented in Table~\ref{tab:impacts}.
To allow for reinterpretation of the data in the context of signal models not considered in this paper, we provide the results of fitting data in the CRs and propagating the prediction to the SRs in \suppMaterialii{} (\ifthenelse{\boolean{cms@external}}{Fig.~18--19 and Tables~3--4}{Fig.~\ref{fig:maskedloose}-\ref{fig:corrtight} and Tables~\ref{tab:maskedloose}-\ref{tab:maskedtight}}).

\section{Summary}

A search is reported for dark matter events with large transverse momentum imbalance and a hadronically decaying top quark.
New t tagging techniques are presented and utilized to identify jets from the Lorentz-boosted top quark.
The data are found to be in agreement with the standard model prediction for the expected background.
Results are interpreted in terms of limits on the production cross section of dark matter (DM) particles via a  flavor-changing neutral current interaction or via the decay of a colored scalar resonance.

Other experimental searches~\cite{monojet13} probe the production of DM via neutral currents, under the assumption that flavor is conserved.
This analysis augments these searches by considering DM production in scenarios that violate flavor conservation.
Assuming $m_\chi = 1\GeV$, $g^\Vrm_\mathrm{u}=0.25$, and $g^\Vrm_{\chi}=1$, spin-1 mediators with masses $0.2 < m_\Vrm<1.75\TeV$ in the FCNC model are excluded at the 95\% confidence level.
Scalar resonances decaying to DM and a top quark are excluded in the range $1.5 < m_\phi < 3.4\TeV$, assuming $m_\psi=100\GeV$.

\begin{acknowledgments}
We congratulate our colleagues in the CERN accelerator departments for the excellent performance of the LHC and thank the technical and administrative staffs at CERN and at other CMS institutes for their contributions to the success of the CMS effort. In addition, we gratefully acknowledge the computing centers and personnel of the Worldwide LHC Computing Grid for delivering so effectively the computing infrastructure essential to our analyses. Finally, we acknowledge the enduring support for the construction and operation of the LHC and the CMS detector provided by the following funding agencies: BMWFW and FWF (Austria); FNRS and FWO (Belgium); CNPq, CAPES, FAPERJ, and FAPESP (Brazil); MES (Bulgaria); CERN; CAS, MoST, and NSFC (China); COLCIENCIAS (Colombia); MSES and CSF (Croatia); RPF (Cyprus); SENESCYT (Ecuador); MoER, ERC IUT, and ERDF (Estonia); Academy of Finland, MEC, and HIP (Finland); CEA and CNRS/IN2P3 (France); BMBF, DFG, and HGF (Germany); GSRT (Greece); OTKA and NIH (Hungary); DAE and DST (India); IPM (Iran); SFI (Ireland); INFN (Italy); MSIP and NRF (Republic of Korea); LAS (Lithuania); MOE and UM (Malaysia); BUAP, CINVESTAV, CONACYT, LNS, SEP, and UASLP-FAI (Mexico); MBIE (New Zealand); PAEC (Pakistan); MSHE and NSC (Poland); FCT (Portugal); JINR (Dubna); MON, RosAtom, RAS, RFBR and RAEP (Russia); MESTD (Serbia); SEIDI, CPAN, PCTI and FEDER (Spain); Swiss Funding Agencies (Switzerland); MST (Taipei); ThEPCenter, IPST, STAR, and NSTDA (Thailand); TUBITAK and TAEK (Turkey); NASU and SFFR (Ukraine); STFC (United Kingdom); DOE and NSF (USA).

\hyphenation{Rachada-pisek} Individuals have received support from the Marie-Curie program and the European Research Council and Horizon 2020 Grant, contract No. 675440 (European Union); the Leventis Foundation; the A. P. Sloan Foundation; the Alexander von Humboldt Foundation; the Belgian Federal Science Policy Office; the Fonds pour la Formation \`a la Recherche dans l'Industrie et dans l'Agriculture (FRIA-Belgium); the Agentschap voor Innovatie door Wetenschap en Technologie (IWT-Belgium); the Ministry of Education, Youth and Sports (MEYS) of the Czech Republic; the Council of Science and Industrial Research, India; the HOMING PLUS program of the Foundation for Polish Science, cofinanced from European Union, Regional Development Fund, the Mobility Plus program of the Ministry of Science and Higher Education, the National Science Center (Poland), contracts Harmonia 2014/14/M/ST2/00428, Opus 2014/13/B/ST2/02543, 2014/15/B/ST2/03998, and 2015/19/B/ST2/02861, Sonata-bis 2012/07/E/ST2/01406; the National Priorities Research Program by Qatar National Research Fund; the Programa Severo Ochoa del Principado de Asturias; the Thalis and Aristeia programs cofinanced by EU-ESF and the Greek NSRF; the Rachadapisek Sompot Fund for Postdoctoral Fellowship, Chulalongkorn University and the Chulalongkorn Academic into Its 2nd Century Project Advancement Project (Thailand); the Welch Foundation, contract C-1845; and the Weston Havens Foundation (USA).

We thank Benjamin Fuks for his help in devising the signal models used in the interpretation of the results.
\end{acknowledgments}

\clearpage
\bibliography{auto_generated}
\ifthenelse{\boolean{cms@external}}{}{
\clearpage
\appendix
\numberwithin{table}{section}
\numberwithin{figure}{section}
\section{Supplementary material\label{sec:supp}}

\begin{figure}[htbp]
  \centering
  \includegraphics[width=0.65\textwidth]{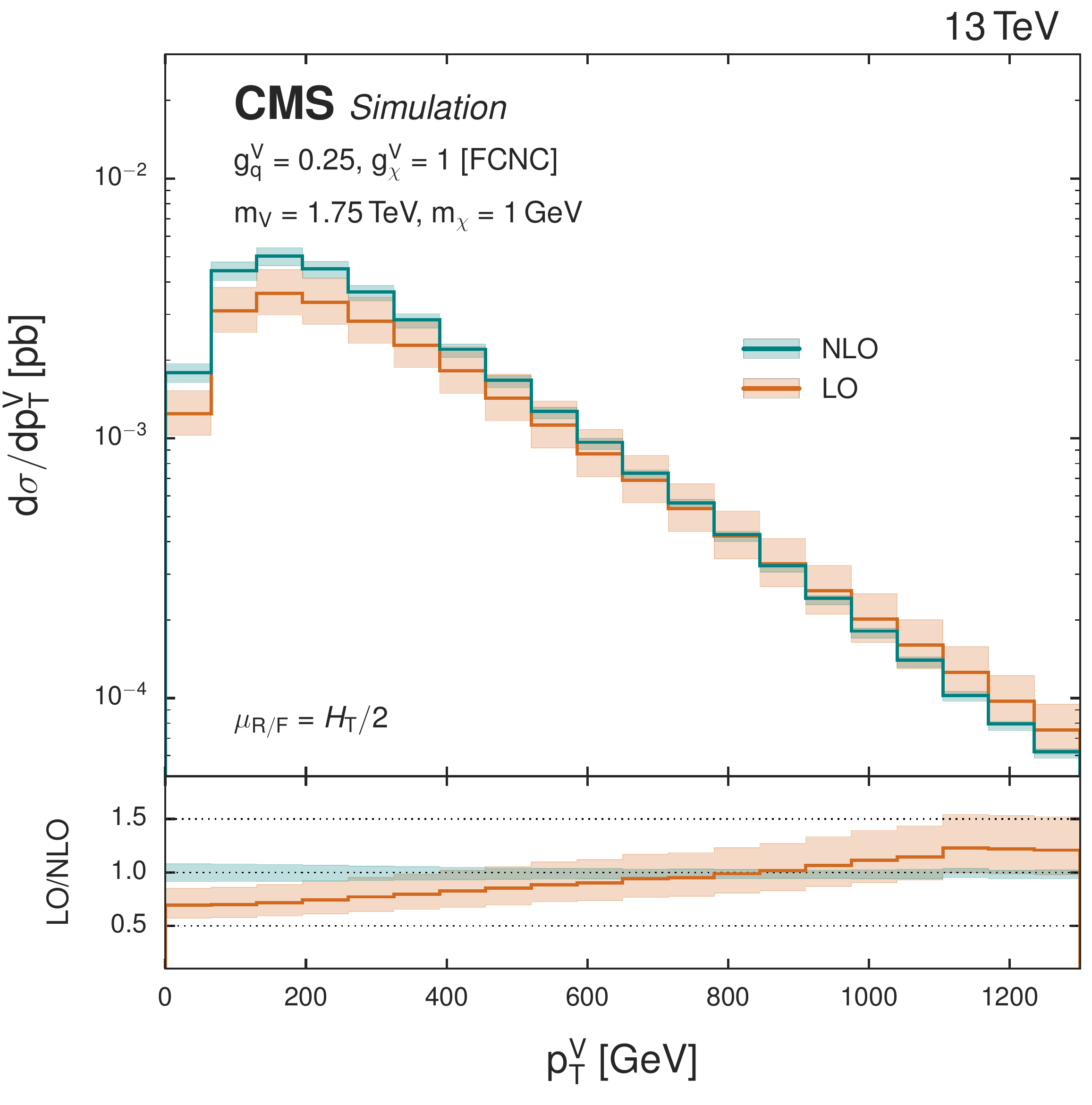}
  \caption{Inclusive distribution of the transverse momentum of the mediator boson V in the FCNC monotop production mechanism, both at leading-order (LO) and next-to-leading order (NLO) accuracy in QCD, assuming couplings of $g_\mathrm{q}^\Vrm = 0.25$ and $g_\chi^\Vrm = 1$ and masses of 1.75\TeV and 1\GeV for V and the fermionic DM particle $\chi$, respectively. Shaded bands around the central predictions correspond to independent variations of the nominal factorization and renormalization scale $H_\mathrm{T}/2$ by factors of 2 and $1/2$. While the NLO case exhibits a softer spectrum for $\pt^\mathrm{V}$ than the LO computation, which should result in a relatively softer $\ptmiss$, the inclusive cross section increases by about 25\% (from 24.8\unit{fb} at LO to 31.4\unit{fb} at NLO). }
    \label{fig:lonlo}
\end{figure}

\begin{figure}[htbp]
  \centering
  \includegraphics[width=0.49\textwidth]{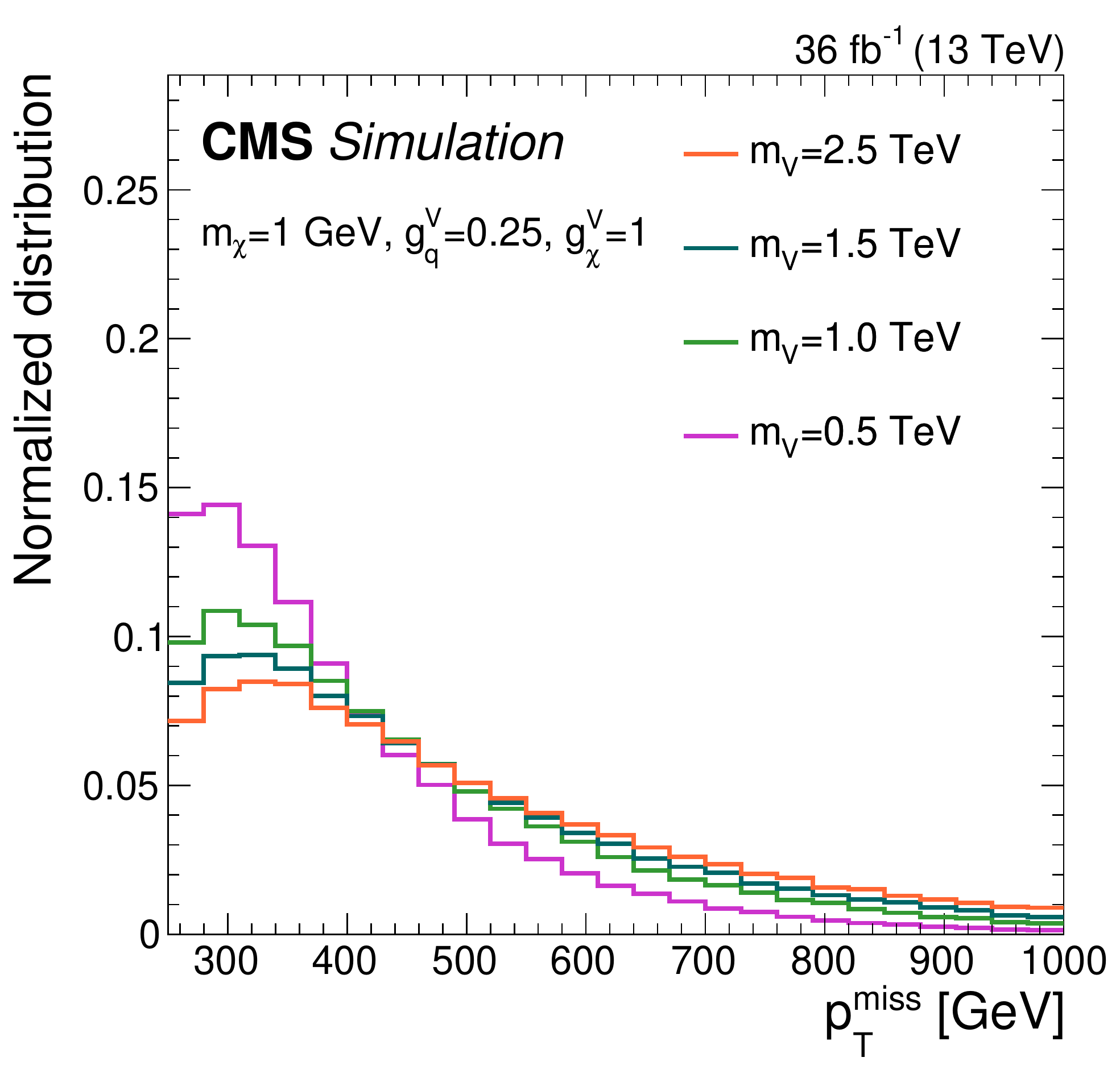}
  \includegraphics[width=0.49\textwidth]{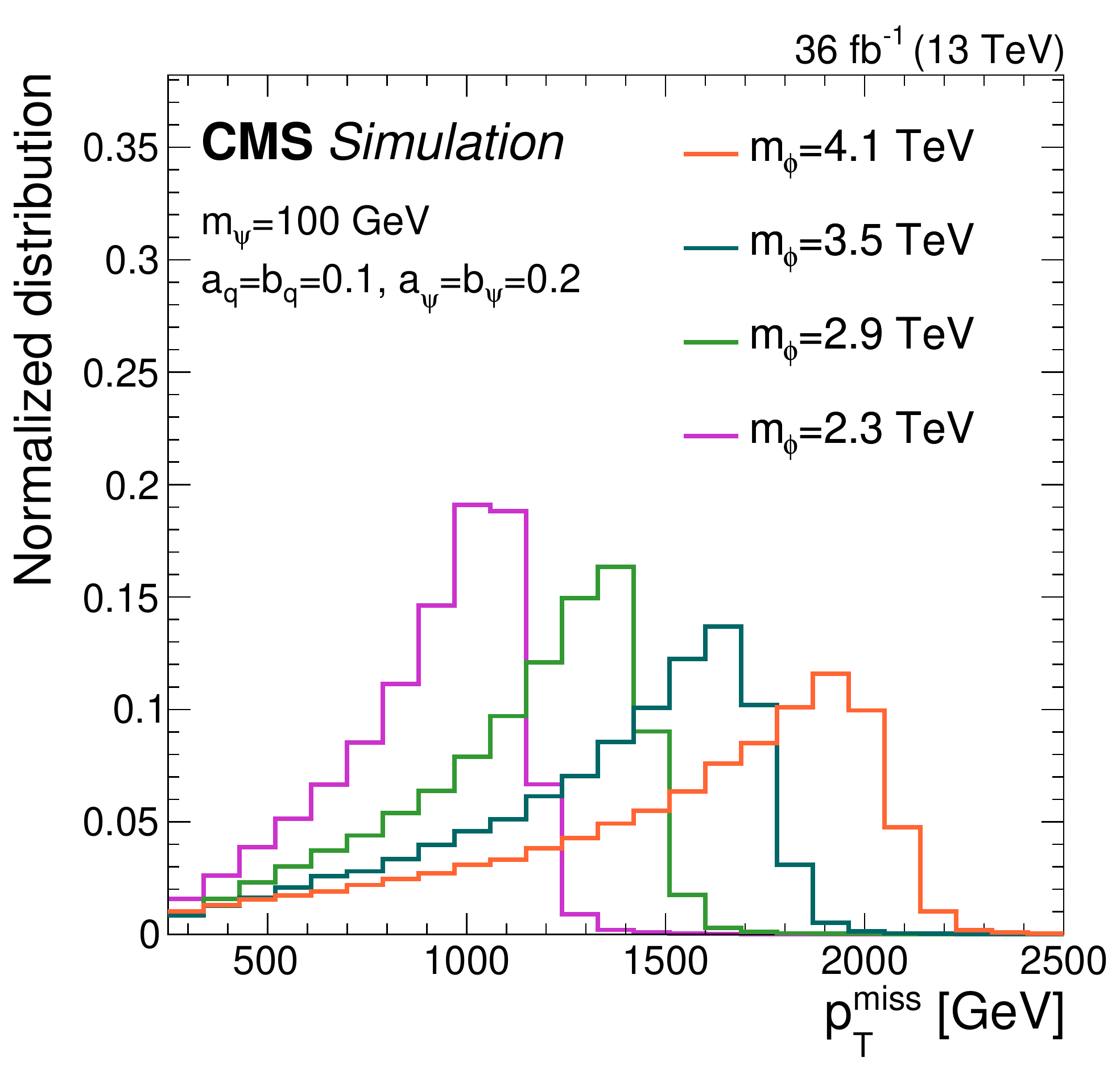}
  \caption{Distribution of \ptmiss in monotop signal models. On the left is shown the FCNC model for various values of $m_\Vrm$; on the right is the scalar resonance model for various values of $m_\phi$ }
  \label{fig:sigkin}
\end{figure}

\begin{table}[htbp]
  \centering
  \topcaption{Importance of groups of systematic uncertainties, as measured by the sensitivity of this search to a benchmark FCNC model ($m_\mathrm{V}=2.25\TeV, m_\chi=1\GeV$). The importance is assessed by evaluating the relative change of the expected 95\% CL limit after removing each group of uncertainties. ``Other sources'' includes all uncertainties not considered elsewhere in the table.}

  \begin{tabular}{cc}
     \hline
       Sources of uncertainty & Change in expected limit (\%) \\ 
     \hline
       Statistical uncertainty in simulation & 3.6 \\ 
       CA15 subjet b tagging   & 1.4 \\ 
       V+jets renorm./fact. scales and PDF & 1.1 \\ 
       Lepton identification   & 0.7 \\ 
       V+jets electroweak corrections & 0.3 \\ 
       V+HF fraction & 0.3  \\ 
       AK4 b tagging   & $<0.1$ \\ 
       Other sources & 0.8 \\ 
     \hline
      \end{tabular}
  \label{tab:impacts}
\end{table}

\begin{figure}[htbp]
  \centering
  \includegraphics[width=0.45\textwidth]{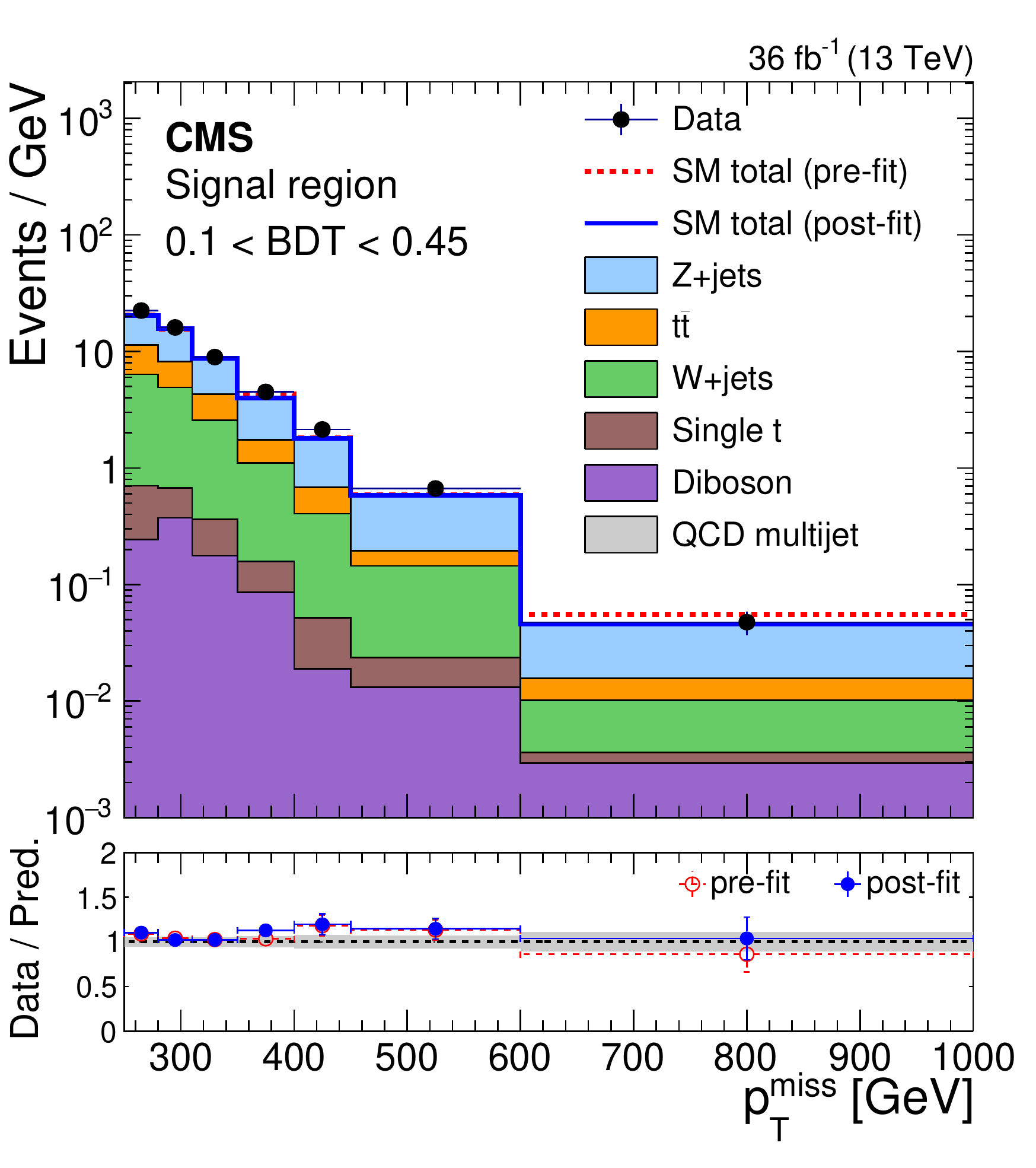}
  \caption{ Distribution of $\ptmiss$ from SM backgrounds and data in the loose category of the signal region after fitting the control regions only.
  Each bin shows the event yields divided by the width of the bin.
  The stacked histograms show the individual SM background distributions after the fit is performed.
      The lower panel of the figure shows the ratio of data to fitted prediction.
      The gray band on the ratio indicates the one standard deviation uncertainty on the prediction after propagating all the systematic uncertainties and their correlations in the fit.
      }
      \label{fig:maskedloose}
\end{figure}

\begin{figure}[htbp]
  \centering
  \includegraphics[width=0.45\textwidth]{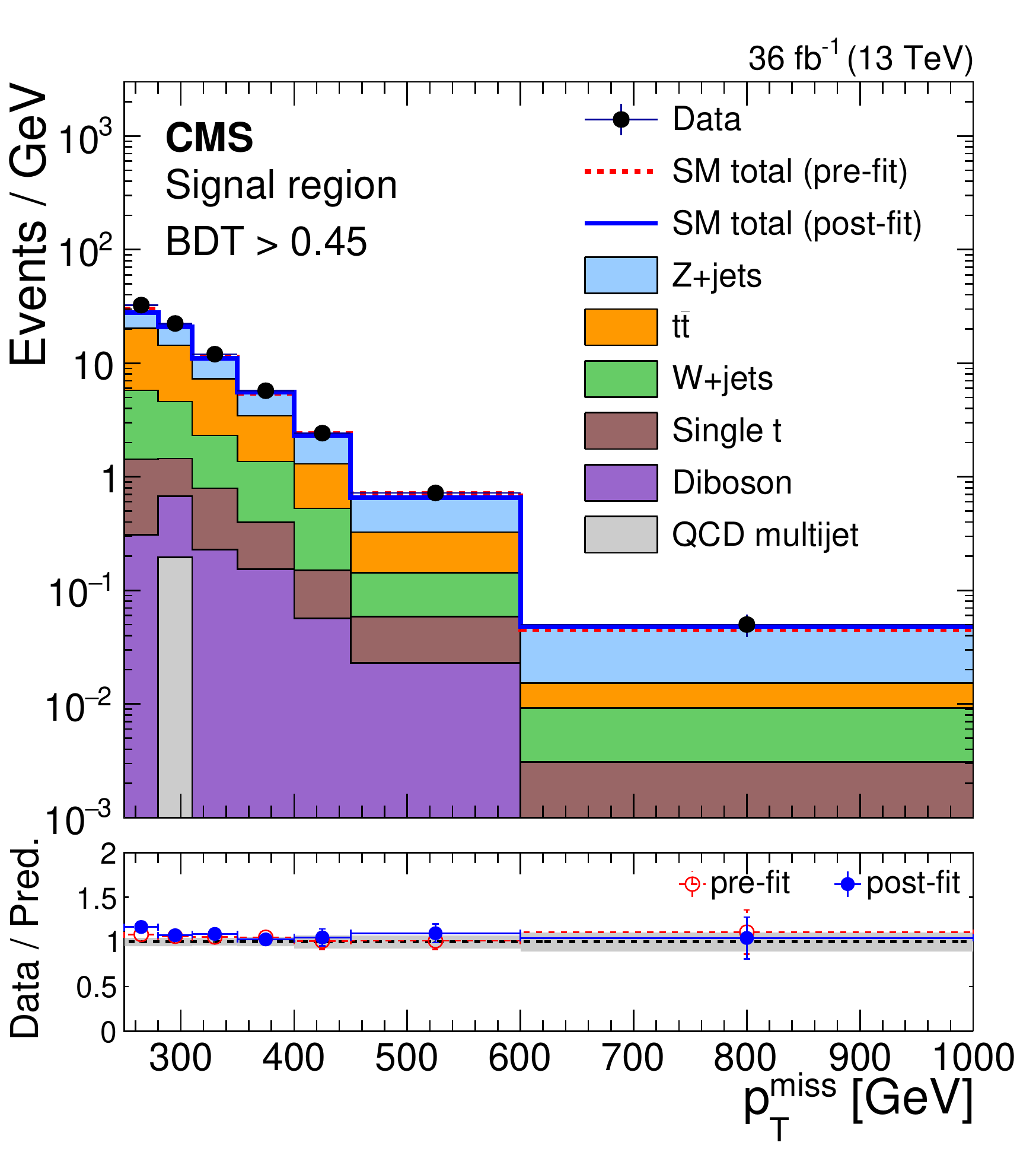}
  \caption{ Distribution of $\ptmiss$ from SM backgrounds and data in the tight category of the signal region after fitting the control regions only.
  Each bin shows the event yields divided by the width of the bin.
  The stacked histograms show the individual SM background distributions after the fit is performed.
      The lower panel of the figure shows the ratio of data to fitted prediction.
      The gray band on the ratio indicates the one standard deviation uncertainty on the prediction after propagating all the systematic uncertainties and their correlations in the fit.
      }
      \label{fig:maskedtight}
\end{figure}

\begin{table}[htbp]
  \centering
  \topcaption{ Predicted SM backgrounds and yields in data in each bin of the loose signal region, after performing the fit in the control regions only. ``Minor backgrounds'' refers to the diboson, single t, and QCD multijet backgrounds. The uncertainties are reported as statistical (driven by the data in the CRs), followed by systematic. }
  \resizebox{\textwidth}{!}{
  \begin{tabular}{lccccccc}
     \hline
       $\ptmiss\,[\GeVns{}]$ & Z+jets  & $\ttbar$ & W+jets  & Minor backgrounds  & Observed & Total backgrounds \\
     \hline
         250--280  & $ 269.8 \pm  5.6 \pm 18.0$ & $ 148.8 \pm  7.1 \pm  7.9$ & $ 170.3 \pm  3.6 \pm 17.6$ & $  21.0 \pm  0.3 \pm  4.1$ & $       673$ & $ 609.9 \pm  9.8 \pm 26.7$\\
         280--310  & $ 226.1 \pm  5.3 \pm 15.0$ & $  98.5 \pm  5.3 \pm  7.3$ & $ 127.2 \pm  3.0 \pm 12.9$ & $  20.2 \pm  0.3 \pm  3.9$ & $       482$ & $ 471.9 \pm  8.1 \pm 21.5$\\
         310--350  & $ 178.4 \pm  4.5 \pm 12.9$ & $  69.1 \pm  4.5 \pm  5.3$ & $  88.2 \pm  2.2 \pm  8.9$ & $  14.5 \pm  0.2 \pm  2.8$ & $       358$ & $ 350.2 \pm  6.8 \pm 16.8$\\
         350--400  & $ 111.9 \pm  3.3 \pm  8.3$ & $  32.4 \pm  3.2 \pm  3.0$ & $  47.1 \pm  1.4 \pm  5.8$ & $   7.9 \pm  0.1 \pm  1.5$ & $       225$ & $ 199.3 \pm  4.8 \pm 10.7$\\
         400--450  & $  55.3 \pm  2.3 \pm  4.4$ & $  13.8 \pm  1.7 \pm  2.1$ & $  17.6 \pm  0.7 \pm  2.3$ & $   2.6 \pm  0.0 \pm  0.5$ & $       107$ & $  89.4 \pm  2.9 \pm  5.4$\\
         450--600  & $  57.9 \pm  2.6 \pm  4.2$ & $   7.6 \pm  1.3 \pm  1.5$ & $  18.1 \pm  0.8 \pm  2.0$ & $   3.5 \pm  0.1 \pm  0.7$ & $       100$ & $  87.2 \pm  3.0 \pm  4.9$\\
        600--1000  & $  12.0 \pm  1.0 \pm  1.2$ & $   2.2 \pm  0.9 \pm  0.8$ & $   2.6 \pm  0.2 \pm  0.4$ & $   1.4 \pm  0.0 \pm  0.3$ & $        19$ & $  18.3 \pm  1.4 \pm  1.5$\\
     \hline
  \end{tabular}

  }
  \label{tab:maskedloose}
\end{table}

\begin{table}[htbp]
  \centering
  \topcaption{ Predicted SM backgrounds and yields in data in each bin of the tight signal region, after performing the fit in the control regions only. ``Minor backgrounds'' refers to the diboson, single t, and QCD multijet backgrounds. The uncertainties are reported as statistical (driven by the data in the CRs), followed by systematic. }
  \resizebox{\textwidth}{!}{

  \begin{tabular}{lccccccc}
     \hline
       $\ptmiss\,[\GeVns{}]$ & Z+jets  & $\ttbar$ & W+jets  & Minor backgrounds  & Observed & Total backgrounds \\
     \hline
         250--280  & $ 224.4 \pm  5.7 \pm 16.9$ & $ 435.9 \pm 10.5 \pm 18.8$ & $ 130.4 \pm  3.3 \pm 15.1$ & $  42.9 \pm  0.8 \pm  9.1$ & $       972$ & $ 833.6 \pm 12.4 \pm 30.9$\\
         280--310  & $ 193.4 \pm  5.8 \pm 16.0$ & $ 293.5 \pm  8.6 \pm 13.7$ & $  94.2 \pm  2.8 \pm 11.5$ & $  37.6 \pm  0.6 \pm  7.2$ & $       671$ & $ 618.6 \pm 10.8 \pm 25.0$\\
         310--350  & $ 149.2 \pm  4.0 \pm 11.0$ & $ 199.1 \pm  6.8 \pm  9.7$ & $  60.6 \pm  1.6 \pm  7.2$ & $  31.7 \pm  0.5 \pm  6.4$ & $       480$ & $ 440.6 \pm  8.1 \pm 17.6$\\
         350--400  & $ 106.1 \pm  4.0 \pm  8.1$ & $ 104.1 \pm  4.6 \pm  5.3$ & $  48.2 \pm  1.8 \pm  5.7$ & $  19.9 \pm  0.3 \pm  3.8$ & $       286$ & $ 278.2 \pm  6.3 \pm 11.9$\\
         400--450  & $  50.2 \pm  2.5 \pm  4.8$ & $  38.6 \pm  2.6 \pm  3.5$ & $  18.9 \pm  0.9 \pm  2.7$ & $   7.5 \pm  0.1 \pm  1.4$ & $       121$ & $ 115.2 \pm  3.7 \pm  6.7$\\
         450--600  & $  49.5 \pm  2.4 \pm  4.8$ & $  27.5 \pm  2.1 \pm  2.8$ & $  12.6 \pm  0.6 \pm  1.9$ & $   8.8 \pm  0.1 \pm  1.7$ & $       108$ & $  98.5 \pm  3.3 \pm  6.1$\\
        600--1000  & $  13.0 \pm  1.3 \pm  1.1$ & $   2.4 \pm  0.5 \pm  0.7$ & $   2.5 \pm  0.2 \pm  0.3$ & $   1.2 \pm  0.0 \pm  0.3$ & $        20$ & $  19.2 \pm  1.4 \pm  1.4$\\
     \hline
      \end{tabular}
  }
  \label{tab:maskedtight}
\end{table}

\begin{figure}[htbp]
  \centering
  \includegraphics[width=0.45\textwidth]{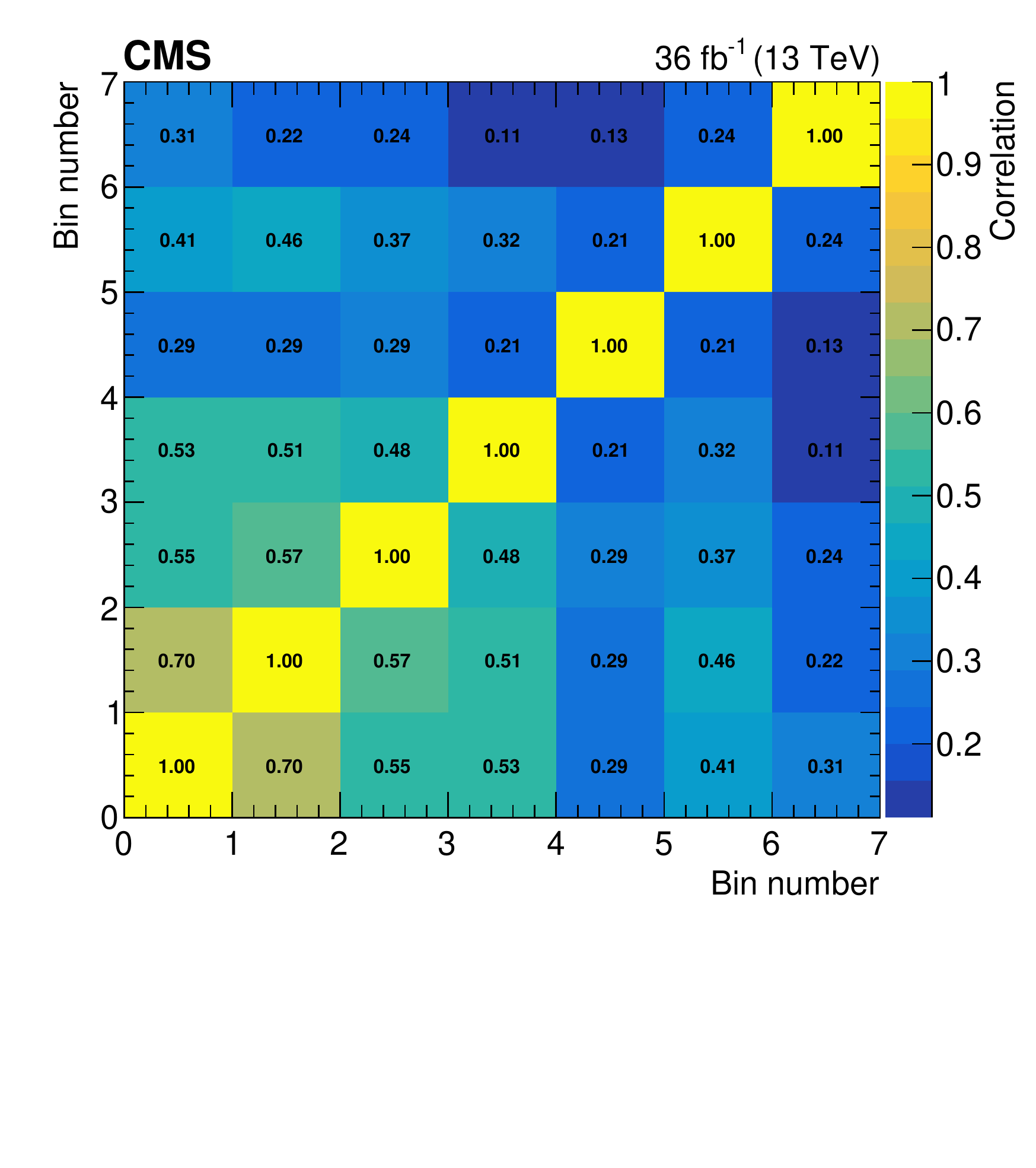}
  \caption{ Correlations between background predictions in each of the bins of the loose signal
  region, after performing the fit in only the control regions.}
      \label{fig:corrloose}
\end{figure}

\begin{figure}[htbp]
  \centering
  \includegraphics[width=0.45\textwidth]{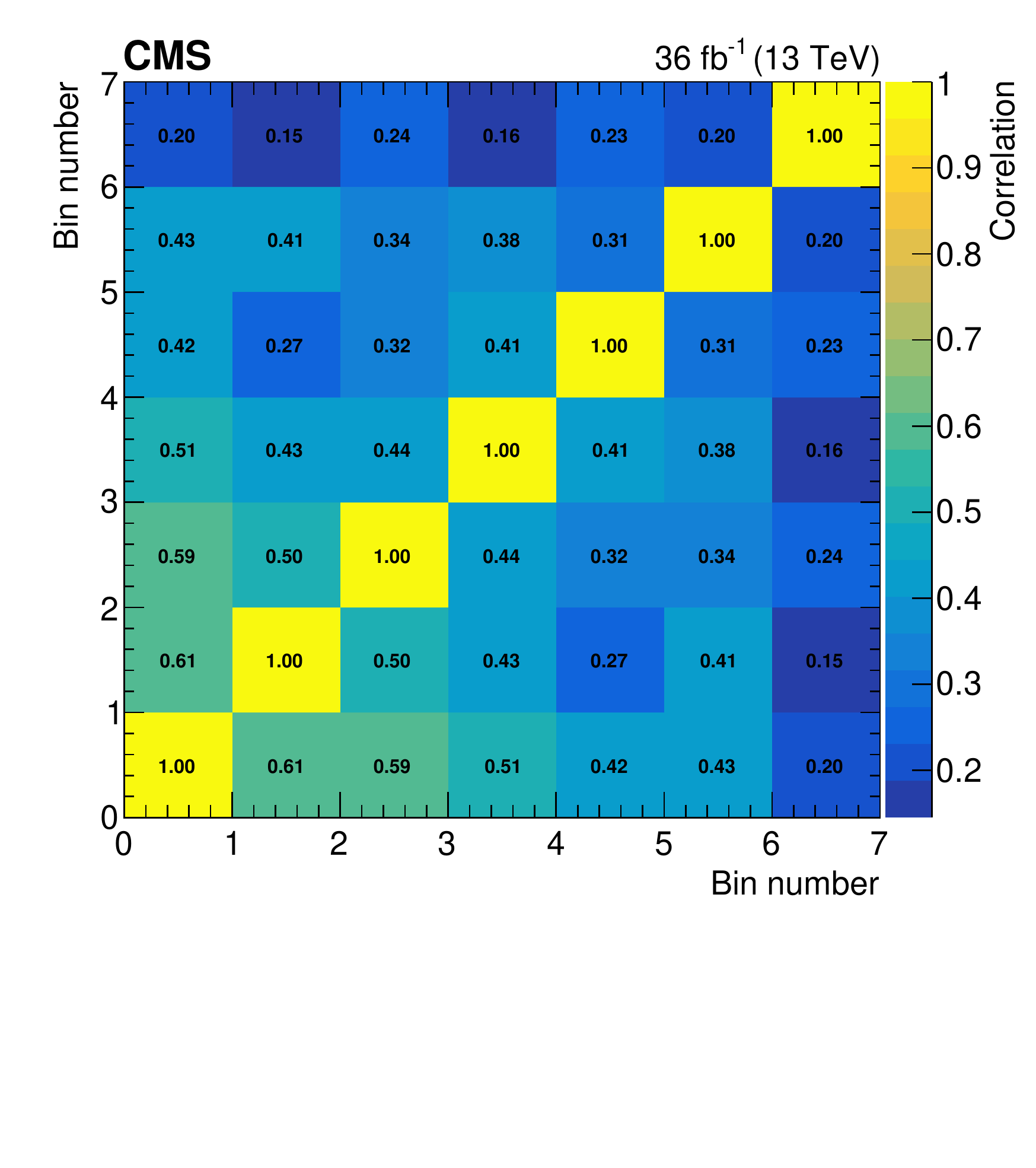}
  \caption{ Correlations between background predictions in each of the bins of the tight signal
  region, after performing the fit in only the control regions.}
      \label{fig:corrtight}
\end{figure}

\begin{figure}[htbp]
  \centering
\includegraphics[width=0.7\textwidth]{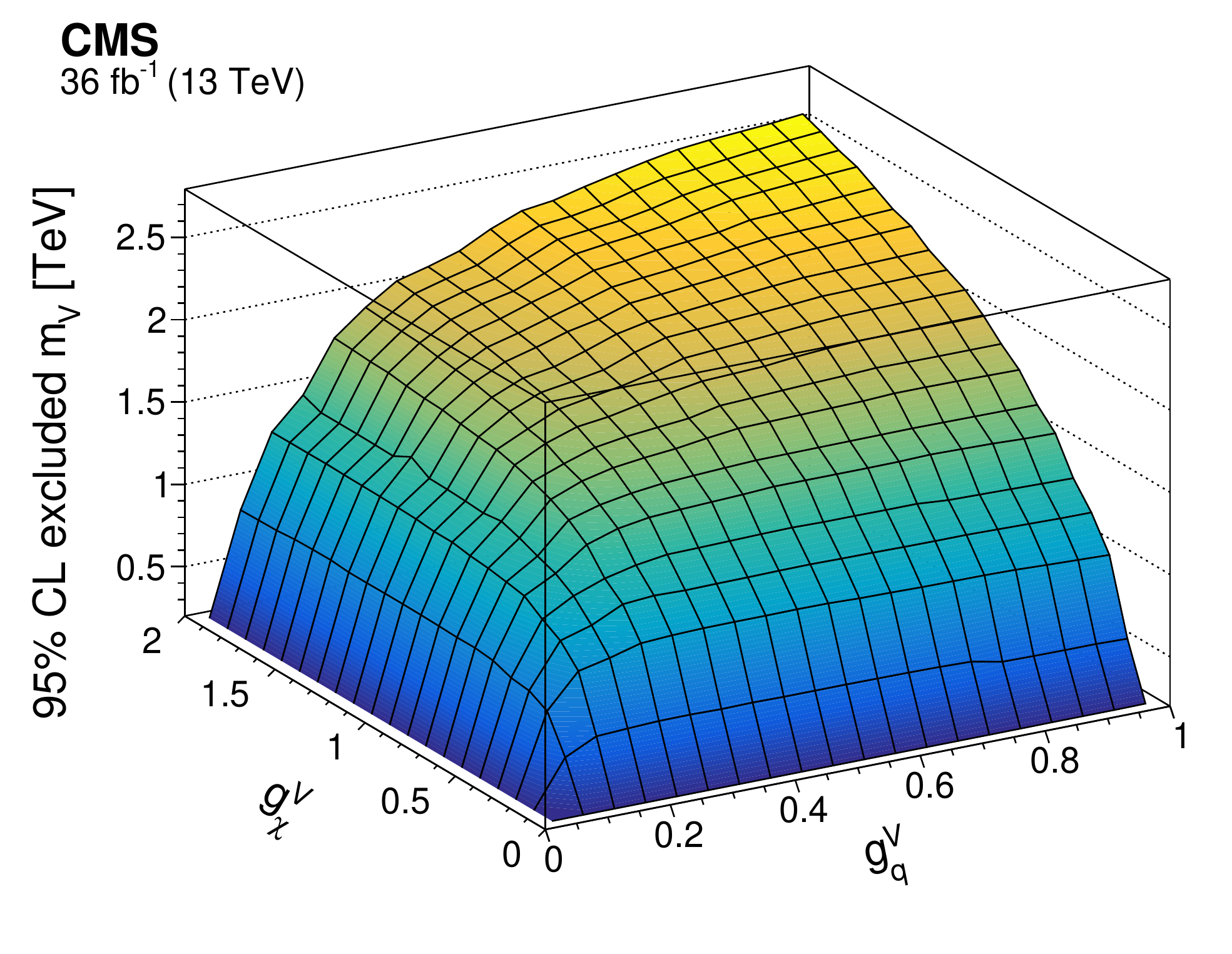}
  \caption{ The maximum excluded mediator mass at 95\% CL as a function of vector couplings to DM and quarks.
  This plot fixes $m_\chi=1\GeV$ and $g_\chi^\mathrm{A} = g_\mathrm{q}^\mathrm{A}= 0$.
  Masses up to 2.5\TeV are excluded given sufficiently large coupling choices.
  }
  \label{fig:3d}
\end{figure}

}
\cleardoublepage \section{The CMS Collaboration \label{app:collab}}\begin{sloppypar}\hyphenpenalty=5000\widowpenalty=500\clubpenalty=5000\vskip\cmsinstskip
\textbf{Yerevan Physics Institute,  Yerevan,  Armenia}\\*[0pt]
A.M.~Sirunyan,  A.~Tumasyan
\vskip\cmsinstskip
\textbf{Institut f\"{u}r Hochenergiephysik,  Wien,  Austria}\\*[0pt]
W.~Adam,  F.~Ambrogi,  E.~Asilar,  T.~Bergauer,  J.~Brandstetter,  E.~Brondolin,  M.~Dragicevic,  J.~Er\"{o},  M.~Flechl,  M.~Friedl,  R.~Fr\"{u}hwirth\cmsAuthorMark{1},  V.M.~Ghete,  J.~Grossmann,  J.~Hrubec,  M.~Jeitler\cmsAuthorMark{1},  A.~K\"{o}nig,  N.~Krammer,  I.~Kr\"{a}tschmer,  D.~Liko,  T.~Madlener,  I.~Mikulec,  E.~Pree,  N.~Rad,  H.~Rohringer,  J.~Schieck\cmsAuthorMark{1},  R.~Sch\"{o}fbeck,  M.~Spanring,  D.~Spitzbart,  W.~Waltenberger,  J.~Wittmann,  C.-E.~Wulz\cmsAuthorMark{1},  M.~Zarucki
% ...
\vskip\cmsinstskip
% ...
1:~Also at Vienna University of Technology,  Vienna,  Austria\\
% ...
\end{sloppypar}
\end{document}